# Thermodynamic Signatures of Phase Separation in Mass-Imbalanced Fermi Mixtures: Superfluid Density of States and Quasiparticle Specific Heat in the $^{163}$Dy–$^{40}$K Atomic Mixture

**N. Ebrahimian**
Department of Physics, Faculty of Basic Sciences, Shahed University, Tehran, 3319118651, Iran
Email: n.ebrahimian@shahed.ac.ir

## Abstract

Mass imbalance changes not only pairing in a polarized Fermi gas but also the thermodynamic condition under which normal and superfluid components can remain phase separated. The low-temperature properties of mass-imbalanced two-component Fermi gases are investigated within the local density approximation on the BCS side of the BCS-BEC crossover, with an unpolarized superfluid spatially separated from a partially polarized normal component. For each interaction strength and mass ratio, the pairing gap, average chemical potential, Hartree-Fock potentials, and imbalance chemical potential are obtained self-consistently, with the latter determined by the normal-superfluid phase separation condition rather than varied independently. The results show that a larger mass ratio requires a larger imbalance chemical potential to maintain phase separation. For the $^{163}$Dy–$^{40}$K mixture, an increase in the absolute value of the interaction strength raises the average and imbalance chemical potentials while reducing the pairing gap. The resulting changes modify the quasiparticle spectrum and the superfluid density of states and are reflected directly in the thermal response, with the quasiparticle specific heat decreasing at a fixed mass ratio. In contrast, increasing the mass ratio enhances the specific heat, which is associated with the larger density of available states near the Fermi surface of the heavier majority component. The results thus connect the normal-superfluid phase separation condition with the quasiparticle spectrum and superfluid density of states, and show that the specific heat provides a clear thermal signature of mass imbalance in the phase-separated regime.

## I. Introduction

Ultracold two-component Fermi gases allow the study of pairing, polarization, and thermodynamic properties with tunable interactions[1-27]. The fundamental concept of a continuous crossover between weak-coupling Bardeen–Cooper–Schrieffer (BCS) pairing and strong-coupling Bose–Einstein condensation (BEC) was first developed theoretically by Eagles and Leggett, showing that the BCS wave function is not limited to weakly interacting systems at zero temperature [3,4], and was later extended to finite temperatures [5]. Further theoretical developments followed, and the crossover was subsequently realized experimentally through magnetic Feshbach resonances [6–27]. The ability to tune the s-wave scattering length makes it possible to investigate the paired state over a wide interaction range while keeping the particle density and population imbalance under independent control [15–27]. Population imbalance introduces an additional competition between pairing and Fermi-surface mismatch. Experiments with ultracold $^{6}$Li gases observed spatial separation between an approximately unpolarized superfluid region and a polarized normal region [1-2,15-16,28-33]. The corresponding first-order normal–superfluid transition is closely related to the Chandrasekhar–Clogston mechanism [34,35]. For a phase-separated state in thermodynamic equilibrium, the chemical potentials of each species are equal in the two phases and the pressures are equal at the interface. These conditions are conveniently implemented by comparing the grand-canonical thermodynamic potentials of the normal and superfluid phases.

Mass imbalance adds a qualitatively distinct control parameter. The unequal-mass problem has been studied in connection with modified phase boundaries, changes in the stability range of the superfluid phase and in the critical temperature, and nontrivial shell structures in traps [36–43]. Experimentally relevant mixtures include $^{40}$K–$^{6}$Li and $^{163}$Dy–$^{40}$K, with mass ratios of approximately 6.7 and 4.08, respectively [44–46]. When a two-component ultracold Fermi gas with s-wave interactions has finite spin polarization, the two spin populations are unequal and their Fermi surfaces become mismatched. This mismatch makes pairing more difficult and can reduce the stability of a uniform paired state. When the energetic cost associated with population imbalance becomes sufficiently large, the system can lower its energy by separating into an unpolarized superfluid and a partially or fully polarized normal phase. The present work considers the partially polarized normal phase**.** For unequal masses, the mass difference also changes the mismatch relevant to pairing and the density of states of the two components, thereby modifying the pairing conditions, the stability of the superfluid state, and the conditions for normal–superfluid phase separation. In trapped ultracold gases, a trapping potential can lead to the spatial arrangement of the phases into a superfluid core and a normal shell.

Ultracold Fermi gases have been extensively studied, including their phase diagrams and thermodynamic properties across different regimes such as normal–superfluid phase separation, as well as critical and transition temperatures [15–16,21–22,30–31,36–39,47–66]. Recent works have investigated the superfluid-normal separation regime on the Bardeen-Cooper-Schrieffer (BCS) side of the BCS–Bose–Einstein condensation (BEC) crossover, focusing on the interface thermal conductivity [32,67-68]. The specific heat has recently become measurable in cold Fermi gas physics [47]. Furthermore, the effect of pairing fluctuations and the pseudogap effect on the specific heat of an ultracold Fermi gas in the BCS-BEC crossover regime was investigated [48]. Previous works on the specific heat include calculations of the total specific heat of spin-polarized, phase-separated Fermi gases with equal masses within Bogoliubov–de Gennes mean-field theory, as well as studies of the specific heat as a function of polarization [50]. In addition, polarized superfluids near their tricritical point and the total specific heat for equal masses were studied [21]. Previous theoretical work investigated the specific heat in the BCS regime without considering normal–superfluid phase separation [22]. However, the combined dependence of the quasiparticle specific heat on mass ratio, interaction strength, and the phase-separation conditions of a mass-imbalanced normal–superfluid system has received considerably less attention.

The present work investigates the quasiparticle specific heat of mass-imbalanced spin-polarized Fermi mixtures in the normal–superfluid phase-separated regime on the BCS side of the crossover. The effects of the imbalance chemical potential, $h_s$ , interaction strength, and mass ratio, $m_r\,(\equiv m_{\downarrow}/m_{\uparrow}$ where $m_{\sigma}$ is the mass of the spin-$\sigma$ population$)$ , are examined. The system is treated within mean-field theory. The local-density approximation (LDA) is used to connect the homogeneous calculation to a trapped gas by treating the trap as a set of locally uniform cells [20-21,37–43,58,64,69-70]. The superfluid component is taken to be unpolarized, whereas the normal component is partially polarized, as illustrated schematically in Fig. 1. The excitation spectrum and the superfluid density of states are also examined, together with the conditions for normal–superfluid phase separation in the unequal-mass system. The calculations are restricted to $h_s < \Delta$, where $\Delta$ denotes the superfluid pairing gap, so that the system remains in the normal–superfluid phase-separated regime rather than entering a Sarma-type state. The $^{163}$Dy–$^{40}$K mixture is used as the principal example, while the mass ratio is varied to show its effect on the calculated quantities. The finite-temperature equations are retained where they define the thermodynamic formulation, and their zero-temperature limits are used for the numerical parameters under the low-temperature conditions discussed later.

The physics of the system is described by the dimensionless interaction strength $1/k_F a$ , which represents the ratio of the characteristic interparticle distance, set by the inverse of the Fermi wave number, $k_F^{-1}$, to the s-wave scattering length, $a$. In the BCS regime, identified by $1/(k_F a) < 0$, the weak attraction leads to the formation of large, overlapping Cooper pairs with a coherence length ξ much larger than the interparticle spacing $(k_F \xi \gg 1)$. Conversely, in the BEC regime, $1/(k_F a) > 0$, the strong attraction results in tightly bound diatomic molecules whose size is much smaller than the average distance between them. It should be noted that the Fermi wave number, $k_F$, is fundamentally defined based on the total particle density of the superfluid component, $n_s \left(\equiv n_{s_\uparrow} + n_{s_\downarrow}\right)$, where $n_{s_\sigma}$ is the particle density of each spin species, through the standard relation $k_F = (3\pi^2 n_s)^{1/3}$.

For unequal masses, the imbalance chemical potential is not fixed by the equal-mass Chandrasekhar–Clogston relation. Instead, the average chemical potential, pairing gap, Hartree–Fock potentials, and imbalance chemical potential are obtained from the coupled gap, number, Hartree–Fock, and phase-separation equilibrium equations. In particular, for each interaction strength and mass ratio, the corresponding value of $h_s$ is determined from equality of the normal and superfluid thermodynamic potentials after the superfluid quantities have been calculated, rather than being varied independently.
The resulting self-consistent quantities are then used in the calculation of the quasiparticle specific heat.

The paper is organized as follows. Section II presents the method and formalism. Sec. II A introduces the Hamiltonian and the superfluid density of states, while Sec. II B gives the gap equation, the number equation, and the thermodynamic potential. Sec. II C presents the quasiparticle specific heat. Sec. III presents the numerical calculations and results, including the phase-separation parameters in Sec. III A, the superfluid density of states in Sec. III B, and the quasiparticle specific heat in Sec. III C. Conclusions are given in Sec. IV.

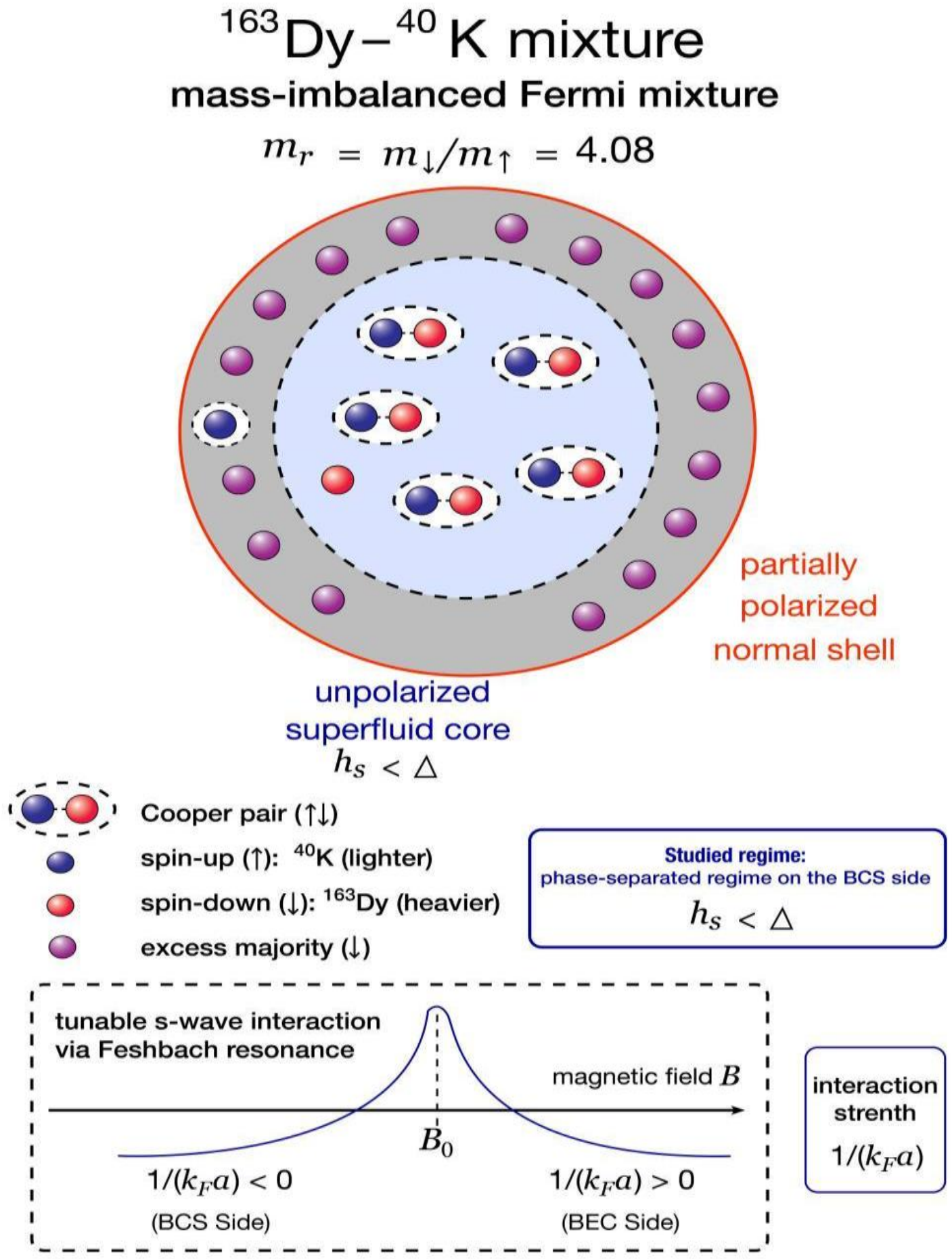

**Fig. 1.** Schematic representation of the $^{163}$Dy–$^{40}$K mass-imbalanced Fermi mixture. In the BCS-side phase-separated regime, an unpolarized superfluid core is surrounded by a partially polarized normal shell. Cooper pairing occurs in the superfluid region, while excess unpaired particles populate the normal shell. The lower schematic illustrates the tuning of the interaction strength $1/k_F a$ through an *s*-wave Feshbach resonance. The calculations are restricted to $h_s < \Delta$ so that other possible phases are avoided in the regime considered here.

## II. Method and Formalism

### A. Hamiltonian and superfluid density of states

A spin-polarized Fermi-Fermi mixture consisting of two fermionic species $\sigma=\uparrow,\downarrow$ with masses $m_i$ and chemical potentials $\mu_\sigma$ is considered. The interspecies interaction is an attractive contact interaction characterized by the coupling constant $g=-4\pi|a|/m_+<0$ where $m_+=2(m_\uparrow^{-1}+m_\downarrow^{-1})^{-1}$.

The effective Hamiltonian of the system may be written as [32,67-68]

$$H_{eff}=\int d^3\mathrm{r}\sum_\sigma[\hat{\psi}_\sigma^\dagger(\mathbf{r})\hat{h}_\sigma(\mathbf{r})\hat{\psi}_\sigma(\mathbf{r})+U_\sigma(\mathbf{r})\hat{\psi}_\sigma^\dagger(\mathbf{r})\hat{\psi}_\sigma(\mathbf{r})$$

$$+\Delta(\mathbf{r})\hat{\psi}_\uparrow^\dagger(\mathbf{r})\hat{\psi}_\downarrow^\dagger(\mathbf{r})+\Delta^*(\mathbf{r})\hat{\psi}_\downarrow(\mathbf{r})\hat{\psi}_\uparrow\hat{\psi}(\mathbf{r})] \quad (1)$$

where $\hat{\psi}_\sigma$ and $\hat{\psi}_\sigma^\dagger$ are fermionic field annihilation and creation operators, respectively; $\Delta(\mathbf{r})$ is the energy gap , and $U_\sigma(\mathbf{r})$ is Hartree-Fock (HF) potential for each spin $\sigma$ in real space and is given by

$$U_\uparrow(\mathbf{r})=g\left\langle\hat{\psi}_\downarrow^\dagger(\mathbf{r})\hat{\psi}_\downarrow(\mathbf{r})\right\rangle\ ,\ U_\downarrow(\mathbf{r})=g\left\langle\hat{\psi}_\uparrow^\dagger(\mathbf{r})\hat{\psi}_\uparrow(\mathbf{r})\right\rangle \quad (2)$$

where $\langle\ldots\rangle$ is the quantum expected value. It should be assumed that there is no magnetized superfluid phase, therefore, the two species of particles with spin up and spin down in the superfluid phase have equal densities. Thus, $U_\uparrow(\mathbf{r})=U_\downarrow(\mathbf{r})\equiv U_s(\mathbf{r})$ where $U_s$ is the HF potential in the superfluid component of the system. The normal phase can be partially polarized, and there exists a phase-separated region where the superfluid and normal states are spatially separated, allowing for the appearance of metastable spin-polarized superfluid solutions. Therefore, in the normal phase, $U_{N\uparrow}(\mathbf{r})\neq U_{N\downarrow}(\mathbf{r})$. For a homogeneous two-component Fermi system at fixed chemical potentials $\mu_\uparrow$ and $\mu_\downarrow$, in the absence of an external potential, the energy gap is assumed to be real and spatially uniform. The single-particle operator $\hat{h}_\sigma(\mathbf{r})$ appearing in Eq. (1) is

$$\hat{h}_\sigma(\mathbf{r})=-\left(\nabla^2/2m_\sigma\right)-\mu_\sigma+V_{trap_\sigma}(\mathbf{r}) \qquad (i=\uparrow,\downarrow) \quad (3)$$

where $V_{trap_\sigma}(\mathbf{r})$ is trapping potential for each spin $\sigma$.

It is crucial to clarify the connection between the theoretical calculations presented in this work for a uniform system and the experimental reality of a Fermi gas confined in a harmonic trap. This connection is rigorously established via the Local Density Approximation (LDA). The application of the LDA is well-justified provided the system contains a sufficiently macroscopic number of particles. Under this assumption, the trapping potential $V_{trap_\sigma}(\mathbf{r})$ for each spin $\sigma$ varies smoothly enough that the inhomogeneous gas can be treated as a continuum of locally uniform cells. The chemical potentials of the spin-$\sigma$ population evaluated at the trap center ($\mu_\sigma$) are thereby replaced by position-dependent local chemical potentials, i.e., $\mu_\sigma(\mathbf{r}) = \mu_\sigma - V_{trap_\sigma}(\mathbf{r})$. Physically, the spatial variation of these local chemical potentials explains the natural emergence of a core-shell geometry. The local average chemical potential, $\mu(\mathbf{r}) = \left(\mu_\uparrow(\mathbf{r}) + \mu_\downarrow(\mathbf{r})\right)/2$, reaches its maximum at the trap center and diminishes toward the edges. The unpolarized superfluid phase is favored at the center of the trap, where the local average chemical potential and the pairing gap are largest. With increasing radial distance, the system reaches the superfluid-normal interface. At this boundary, the pairing energy of the superfluid is balanced by the energy associated with the population imbalance in the normal phase. Thus, the theoretical calculations performed for a uniform system describing a fully unpolarized superfluid directly map onto the local state of matter within the central core of the trap, where the excess particles (or residual population imbalance) are entirely expelled to the surrounding normal shell.

When considering LDA and the role of trapping potentials in mass-imbalanced systems, the trapping potentials for the two species can be unequal $V_{trap_\uparrow}(\mathbf{r}) \neq V_{trap_\downarrow}(\mathbf{r})$. This inequality alters the local imbalance chemical potential, given by $h(\mathbf{r}) = \mu_\uparrow(\mathbf{r}) - \mu_\downarrow(\mathbf{r}) = h_s - [V_{trap\uparrow}(\mathbf{r}) - V_{trap\downarrow}(\mathbf{r})]/2$ where the imbalance chemical potential, $h_s$, is defined by $\left(\mu_\uparrow - \mu_\downarrow\right)/2$ and is independent of $\mathbf{r}$. The two components can therefore experience different spatial confinement, resulting in different density profiles. This difference in their density profiles reduces the spatial overlap region in which the two species interact. If $h(\mathbf{r})$ away from center exceeds the critical limit required to support s-wave pairing, the spatial extent of the superfluid core can shrink compared to the equal-mass case. However, in the numerical calculations of different mass ratios, the same trapping potential is assumed for both species; therefore, the effects of the trap can be included in the local average chemical potential, while the term $h(\mathbf{r})$ remains fixed and equals $h_s$ [32].

When field creation and annihilation operators, $\hat{\psi}^\dagger_\sigma(\mathbf{r})$ and $\hat{\psi}_\sigma(\mathbf{r})$ are written in terms of fermionic creation and annihilation operators, $a^\dagger_{\mathbf{k},\sigma}$ and $a_{\mathbf{k},\sigma}$ with momentum vector $\mathbf{k}$ and spin $\sigma$, as $\hat{\psi}^\dagger_\sigma(\mathbf{r}) = \sum_{\vec{k}} e^{i\mathbf{k}.\mathbf{r}} a^\dagger_{\mathbf{k},\sigma}$ and $\hat{\psi}_\sigma(\mathbf{r}) = \sum_{\mathbf{k}} e^{i\mathbf{k}.\mathbf{r}} a_{\mathbf{k},\sigma}$, in momentum space the effective Hamiltonian becomes (throughout the paper, $k_B = \hbar = 1$ is considered)

$$H_{eff} = \sum_{\mathbf{k},\sigma}\left(\frac{k^2}{2m} - \mu_\sigma + U_\sigma\right) a^\dagger_{\mathbf{k},\sigma} a_{\mathbf{k},\sigma} - \sum_{\mathbf{k}} \Delta\left(a^\dagger_{\mathbf{k},\uparrow} a^\dagger_{-\mathbf{k},\downarrow} + a_{-\mathbf{k},\downarrow} a_{\mathbf{k},\uparrow}\right) \tag{4}$$

where $U_\sigma$ is the Fourier transform of $U_\sigma(\mathbf{r})$ in momentum space. Using the inverse of the Bogoliubov transformation,

$$\begin{pmatrix} \gamma_{\mathbf{k}\alpha} \\ \gamma^\dagger_{\mathbf{k}\beta} \end{pmatrix} = \begin{pmatrix} u^*_{\mathbf{k}} & -v_{\mathbf{k}} \\ v^*_{\mathbf{k}} & u_{\mathbf{k}} \end{pmatrix} \begin{pmatrix} a_{\mathbf{k}\uparrow} \\ a^\dagger_{-\mathbf{k}\downarrow} \end{pmatrix} \tag{5}$$

the effective Hamiltonian can be written in terms of Bogoliubov operators, $\gamma^\dagger$ and $\gamma$ and is thereby diagonalized. The Bogoliubov transformation presented in Eq. (5) is strictly consistent with the definitions of the $\alpha$ and $\beta$ channels. This means that Eq. (5), which yields the transformation for $\gamma^\dagger_{\vec{k}\alpha}$ ($\gamma^\dagger_{\mathbf{k}\beta}$), indicates that the $\alpha$ ($\beta$) channel consists of a combination of a spin-up quasiparticle and a spin-down quasihole (vice versa). As is standard in quantum many-body physics, the original particle operators ($a^\dagger$ and $a$) obey the fermionic anticommutation relations, i.e., $\left\{a_{\mathbf{k}\sigma}, a^\dagger_{\mathbf{k}'\sigma'}\right\} = \delta_{\mathbf{k}\mathbf{k}'}\delta_{\sigma\sigma'}$, and $\left\{a_{\mathbf{k}\sigma}, a_{\mathbf{k}'\sigma'}\right\} = \left\{a^\dagger_{\mathbf{k}\sigma}, a^\dagger_{\mathbf{k}'\sigma'}\right\} = 0$. By assuming the coefficients $u_k$ and $v_k$ to be real and taking into account their normalization condition, i.e., $u_{\mathbf{k}}{}^2 + v_{\mathbf{k}}{}^2 = 1$, the transformation remains canonical. This normalization mathematically guarantees that the new Bogoliubov operators ($\gamma$ and $\gamma^\dagger$) also perfectly satisfy the fundamental fermionic anti-commutation relations, such as( for $\alpha$ channel), $\{\gamma_{\mathbf{k}\alpha}, \gamma_{\mathbf{k}'\alpha}\} = 0,\ \ \{\gamma^\dagger_{\mathbf{k}\alpha}, \gamma^\dagger_{\mathbf{k}'\alpha}\} = 0$ and $\{\gamma_{\mathbf{k}\alpha}, \gamma^\dagger_{\mathbf{k}'\alpha}\} = \delta_{\mathbf{k}\mathbf{k}'}$. Finally, the diagonalized Hamiltonian then takes the following form in terms of these Bogoliubov operators (see Appendix A for details).

$$H_{diag} = \sum_{\mathbf{k}} \left( \frac{k^2}{2m_+} - \mu_s - \sqrt{\left(\frac{k^2}{2m_+} - \mu_s\right)^2 + \Delta^2} \right) + \sum_{\mathbf{k}} \left( E_{\mathbf{k}_\alpha} \gamma^\dagger_{\mathbf{k}\alpha}\gamma_{\mathbf{k}\alpha} + E_{\mathbf{k}_\beta} \gamma^\dagger_{\mathbf{k}\beta}\gamma_{\mathbf{k}\beta} \right) - \frac{\Delta^2}{g} \tag{6}$$

where the first summation and the last term in Eq. (6) were obtained within the mean-field approximation (see Appendix A for details). Also, $E_{\vec{k}_\alpha}$ and $E_{\vec{k}_\beta}$, which are the excitation energies of the $\alpha$ and $\beta$ channels in the superfluid component, respectively, are given by

$$E_{\mathbf{k}_\alpha} \equiv -h_s + \frac{k^2}{2m_-} + \sqrt{\left(\frac{k^2}{2m_+} - \mu_s\right)^2 + \Delta^2}\ \ ,\ \ E_{\mathbf{k}_\beta} \equiv h - \frac{k^2}{2m_-} + \sqrt{\left(\frac{k^2}{2m_+} - \mu_s\right)^2 + \Delta^2} \tag{7}$$

where $m_\pm = 2\left(m_\uparrow^{-1} \pm m_\downarrow^{-1}\right)^{-1}$. When the HF potential is included, the average chemical potential, $\mu_s$, is defined as $\mu_s = \left((\mu_\uparrow + \mu_\downarrow)/2\right) - U_s$. Unlike the previously defined imbalance chemical potential, $h_s = (\mu_\uparrow - \mu_\downarrow)/2$ (assuming that $h_s > 0$), $\mu_s$ contains the $U_s$ term. Both channels are included in the diagonalized Hamiltonian and in the partition function. Numerical evaluation of the $\beta$ channel contribution shows essentially no change in the numerical results presented; the $\alpha$ channel is therefore used for the density of states (DOS) and superfluid specific heat. It should be noted that the energy spectra of the polarized Fermi gas (Eq. (7)) are very different from those of the conventional Fermi gas due to the existence of the mass ratio and the self-consistently determined imbalance and average chemical potentials. Therefore, the physical quantities of the polarized Fermi gas depend on $1/k_F a$, $\mu_s$ and $h_s$, so that all physical quantities of this system differ from those of the conventional Fermi gas. All relevant thermodynamic quantities (such as the order parameter $\Delta$, average chemical potential, $\mu_s$, imbalance chemical potential, $h_s$, and temperature $T$) by $E_F$ are scaled. The physical justification for this approach is that the universal behavior of the system can be extracted. By normalizing the parameters, the results become independent of the absolute density of a specific experiment. This scaling allows the results to be compared for systems characterized by the same dimensionless parameters. The reference Fermi energy is defined as $E_F = k_F^2/2m_+$. By defining $k_F$ based on a fixed reference total density $n_s$, $E_F$ and $k_F^{-1}$ are established as the constant units of energy and length, respectively.

Also, the superfluid density of states is obtained through $DOS = \sum_n \delta(E - E_n)$, where $\delta(E - E_n)$ is the delta function. For the $\alpha$ channel the resulting expression can be written as

$$DOS = \frac{4\pi k(E)}{\frac{1}{m_-} + \frac{\left(\frac{1}{m_+}\right)\left(\frac{k^2}{2m_+} - \mu_s\right)}{\sqrt{\left(\frac{k^2}{2m_+} - \mu_s\right)^2 + \Delta^2}}} \tag{8}$$

where

$$k^2(E) = \frac{1}{\left(\frac{1}{(2m_-)^2} - \frac{1}{(2m_+)^2}\right)} \left( -\left[\frac{\mu_s}{2m_+} - \frac{E_{\vec{k}_\alpha} + h_s}{2m_-}\right] + \sqrt{\left[\frac{\mu_s}{2m_+} - \frac{E_{\vec{k}_\alpha} + h_s}{2m_-}\right]^2 - \left(\frac{1}{(2m_-)^2} - \frac{1}{(2m_+)^2}\right)\left(\left(E_{\vec{k}_\alpha} + h_s\right)^2 - \Delta^2 - \mu_s^2\right)} \right) \tag{9}$$

Eqs. (8) and (9) explicitly show the dependence of the superfluid density of states on the mass ratio, the gap, and the average and imbalance chemical potentials.

### B. The gap equation, the number equation, and the thermodynamic potential

The gap and number equations determine the self-consistent quantities of the superfluid component. At finite temperature the gap equation obtained from the Bogoliubov transformation is

$$\begin{aligned}\Delta &= -g\sum_{\vec{k}} \langle a_{-\mathbf{k}\downarrow} a_{\mathbf{k}\uparrow} \rangle = -g\sum_{\mathbf{k}} -v_{\mathbf{k}} u_{\mathbf{k}} \langle \gamma_{\mathbf{k}\alpha}^{\dagger} \gamma_{\mathbf{k}\alpha} \rangle + v_{\mathbf{k}} u_{\mathbf{k}} \langle 1 - \gamma_{\mathbf{k}\beta}^{\dagger} \gamma_{\mathbf{k}\beta} \rangle \\ &= -g\sum_{\mathbf{k}} v_{\mathbf{k}} u_{\mathbf{k}} \left(1 - f\left(E_{\mathbf{k}_\alpha}\right) - f\left(E_{\mathbf{k}_\beta}\right)\right)\end{aligned} \tag{10}$$

where $\langle \gamma_{\mathbf{k}\alpha}^{\dagger} \gamma_{\mathbf{k}\alpha} \rangle = f\left(E_{\mathbf{k}_\alpha}\right) \equiv 1/(1 + e^{E_{\mathbf{k}_\alpha}/T})$ and $\langle \gamma_{\mathbf{k}\beta}^{\dagger} \gamma_{\mathbf{k}\beta} \rangle = f\left(E_{\mathbf{k}_\beta}\right) \equiv 1/(1 + e^{E_{\mathbf{k}_\beta}/T})$ in which $f$ is the Fermi-Dirac distribution function. To remove the divergence in the gap equation, Eq. (10), the relation between the contact interaction and the scattering length, as given in Ref. [71], is used. With $m_+$ replacing $m$, the relation is

$$\frac{-m_+ g}{4\pi a}+1=-\frac{g}{2}\int\frac{d^3k}{(2\pi)^3}\left(\frac{1}{\frac{k^2}{2m_+}}\right) \tag{11}$$

Then, by substituting $u_k v_k$ given by Eq. (A5) into Eq. (10), and using Eq. (11) in the gap equation gives the finite-temperature form

$$1=-\frac{g}{2}\int\frac{d^3k}{(2\pi)^3}\left(\frac{1-f\left(E_{\mathbf{k}_\alpha}\right)-f\left(E_{\mathbf{k}_\beta}\right)}{\sqrt{\left(\frac{k^2}{2m_+}-\mu_s\right)^2+\Delta^2}}-\frac{1}{\frac{k^2}{2m_+}}\right) \tag{12}$$

In the low-temperature regime used for the numerical results, the Fermi-Dirac distribution functions in Eq. (12) are negligible in the superfluid component, and the gap equation reduces to

$$1=-\frac{g}{2}\int\frac{d^3k}{(2\pi)^3}\left(\frac{1}{\sqrt{\left(\frac{k^2}{2m_+}-\mu_s\right)^2+\Delta^2}}-\frac{1}{\frac{k^2}{2m_+}}\right) \tag{13}$$

In dimensional regularization[72], the second term in the integrand of Eq. (13) can be dropped. From Eqs. (B1), (B2) and (B6), and using $k_F\equiv\left(3\pi n_s\left(T=0\right)\right)^{1/3}$ , the zero-temperature gap and number equations can be reduced to the following relations which are used to determine $\mu_s$ and $\Delta$

$$k_F^3=-\frac{9\pi^2}{2}\sqrt{2}m_+^{3/2}\mu_s^{3/2}\left(\pi\left(1+X^2\right)^{3/4}p_{3/2}\left(-\frac{1}{\sqrt{1+X^2}}\right)+\pi\left(1+X^2\right)^{1/4}p_{1/2}\left(-\frac{1}{\sqrt{1+X^2}}\right)\right) \tag{14}$$

$$k_F a=\left(3\pi/4\right)^{1/3}\left[\left(P_{3/2}\left(-\frac{1}{\sqrt{1+X^2}}\right)/P_{1/2}\left(-\frac{1}{\sqrt{1+X^2}}\right)+\left(\frac{1}{\sqrt{1+X^2}}\right)\right)\Big/P_{1/2}^2\left(-\frac{1}{\sqrt{1+X^2}}\right)\right]^{1/3} \tag{15}$$

where $X\equiv\mu_s/\Delta$. For a given interaction strength, $1/k_F a$, Eq. (15) determines $X$. Using $k_F=\sqrt{2m_+T_F}$ and the normalization $E_F=k_BT_F=1$, where $E_F$ and $T_F$ are Fermi energy and Fermi temperature, respectively, Eq. (14) then determines $\mu_s$ and the gap is obtained from $\Delta\equiv\mu_s/X$. The HF potential in the superfluid component, $U_s$, is obtained from the number density in the superfluid component through $U_s=gn_s\left(T=0\right)=\left(4\pi a/m_+\right)n_s\left(T=0\right)$. Using Eq. (B1), this relation can then be written in terms of $\mu_s$ and $X\left(\equiv\mu_s/\Delta\right)$ as

$$U_s = \mu_s \left( 1 + \sqrt{1+x^2} P_{3/2} \left( \frac{-1}{\sqrt{1+x^2}} \right) / P_{1/2} \left( \frac{-1}{\sqrt{1+x^2}} \right) \right) \tag{16}$$

Therefore, given $X$ and $\mu_s$, $U_s$ can be obtained numerically from Eq. (16). This value of $U_s$ is then used in the condition for normal–superfluid phase separation, which is discussed later. Furthermore, the number equation for the normal component is required to calculate the specific heat of this component. For the normal component, the number density of spin $\sigma$ with mass $m_\sigma$ at finite temperature is

$$n_{N_\sigma}(T) = \int \frac{d^3k}{(2\pi)^3} f\left( \frac{k_F^2}{2m_\sigma} - \mu_\sigma + U_{N_\sigma} \right) \tag{17}$$

where the index $N$ denotes the normal case, and the Fermi-Dirac distribution function, $f$, depends on $k_F^2/2m_\sigma - \mu_\sigma + U_{N_\sigma}$ where $U_{N_\sigma}$ denotes the HF potential in normal component. Then, for the normal component, the result is

$$n_{N_\sigma}(T) = \frac{(2m_\sigma)^{3/2}}{4\pi^2} T^{3/2} \Gamma(3/2) F_{1/2}\left( \left( \mu_\sigma - U_{N_\sigma} \right) / T \right) \tag{18}$$

where $\Gamma$ is the gamma function and $F$ is given by the following relation

$$F_j(\eta) = \frac{1}{\Gamma(j+1)} \int_0^\infty dx \frac{x^j}{e^{x-\eta}+1} \tag{19}$$

and its zero-temperature limit is

$$n_{N_\sigma} = \int_{k \le k_{F_\sigma} = \sqrt{2m_\sigma(\mu_\sigma - U_{N_\sigma})}} \frac{d^3k}{(2\pi)^3} = \left. \frac{k_{F,\sigma}^3}{6\pi^2} \right|_{k_{F_\sigma} = \sqrt{2m_\sigma(\mu_\sigma - U_{N_\sigma})}} = \frac{\left[ 2m_\sigma \left( \mu_\sigma - U_{N_\sigma} \right) \right]^{3/2}}{6\pi^2} \tag{20}$$

where $k_{F_\sigma}$ is Fermi wave number for spin $\sigma$. Furthermore, since $U_{N_\sigma} = g n_{N_{-\sigma}}$, using $\mu_\downarrow = \mu_s + U_s - h_s$, and $g = 4\pi a / m_+$, the spin-up HF potential is obtained as

$$U_{N_\uparrow} = g n_{N_\downarrow} = \frac{2a}{m_+} \frac{(2m_r)^{3/2}}{3\pi} \left( \mu_s + U_s - h_s - U_{N_\downarrow} \right)^{3/2} \tag{21}$$

Using $\mu_\uparrow = \mu_s + U_s + h_s$, the spin-down HF potential is

$$U_{N_\downarrow} = \frac{4a}{m_+} \frac{(2)^{1/2}}{3\pi} \left( \mu_s + U_s + h_s - U_{N_\uparrow} \right)^{3/2} \tag{22}$$

where $m_\uparrow = 1$ is used. Eqs. (14)-(16) and (21)-(22) determine $\mu_s$, $\Delta$, $U_s$, $U_{N_\uparrow}$, and $U_{N_\downarrow}$. To obtain $h_s$, the equilibrium condition for phase separation is imposed by comparing the grand-canonical thermodynamic potentials of the normal and superfluid phases. For the superfluid component, $Z = Tr e^{-H_{diag}/T}$ is the partition function, where $Tr$ denotes the trace over the Hilbert (Fock) space, and $\Omega_s = -T\ln Z$. Using the diagonal Hamiltonian, one obtains(Appendix C)

$$\Omega_s = \frac{\Delta^2}{g} + \sum_k \left( k^2/2m_+ - \mu_s - \sqrt{\left(k^2/2m_+ - \mu_s\right)^2 + \Delta^2} \right) - T\sum_k \left( \ln\left(1+e^{-E_{k\alpha}/T}\right) + \ln\left(1+e^{-E_{k\beta}/T}\right) \right) \quad (23)$$

The first two terms in Eq. (23) arise from the mean-field part of the diagonalized Hamiltonian, while the logarithmic terms account for the thermal contributions from the $\alpha$ and $\beta$ channels. In the zero-temperature limit, $\Omega_s$ reduces to the following expression (Appendix C)

$$\Omega_s = -\frac{\sqrt{2}}{20\pi}\mu_s^{5/2} m_+^{3/2}\left(1+X^2\right)^{1/4}\left[-4\sqrt{1+X^2}P_{3/2}\left(-\left(1+X^2\right)^{-1/2}\right) + \left\{\left(1+X^2\right)-5\right\}P_{1/2}\left(-\left(1+X^2\right)^{-1/2}\right)\right] \quad (24)$$

Hamiltonian of the normal component is given by

$$H_N = \sum_{k,\sigma}\left(\frac{k^2}{2m_\sigma} - \mu_\sigma + U_{N_\sigma}\right) a_{k\sigma}^\dagger a_{k\sigma} \quad (25)$$

where $U_{N_\sigma}$ is HF potential in the normal component. Also, the partition function becomes

$$Z_N = Tre^{-H_N/T} = \prod_{k,\sigma}\left(1+e^{-\left(\frac{k^2}{2m_\sigma}-\mu_\sigma+U_{N_\sigma}\right)/T}\right) \quad (26)$$

and the grand-canonical thermodynamic potential is given by

$$\Omega_N(T) = -T\ln Z_N = -T\sum_{\sigma=\uparrow,\downarrow}\sum_{\vec{k}} \ln\left(1+e^{-\left(\frac{k^2}{2m_\sigma}-\mu_\sigma+U_{N_\sigma}\right)/T}\right) \quad (27)$$

The sum in Eq. (27) runs over the two spin species, and each term contains the corresponding mass, chemical potential, and HF potential in normal component. Mass and population imbalance therefore enter $\Omega_N$ through the spin-dependent dispersions. Finally, at zero temperature, the total thermodynamic potential for normal component becomes (Appendix C)

$$\begin{aligned}\Omega_N = \Omega_{N_\uparrow} + \Omega_{N_\downarrow} &= -\frac{\left(2m_\uparrow\right)^{3/2}}{15\pi^2}\left(\mu_\uparrow - U_{N_\uparrow}\right)^{5/2} - \frac{\left(2m_\downarrow\right)^{3/2}}{15\pi^2}\left(\mu_\downarrow - U_{N_\downarrow}\right)^{5/2} \\ &= -\frac{\left(2\right)^{3/2}}{15\pi^2}\left(\mu_s + U_s + h_s - U_{N_\uparrow}\right)^{5/2} - \frac{\left(2m_r\right)^{3/2}}{15\pi^2}\left(\mu_s + U_s - h_s - U_{N_\downarrow}\right)^{5/2}\end{aligned} \quad (28)$$

To establish normal–superfluid phase separation, three conditions must be satisfied. First, the chemical potentials of each spin species are equal in the normal and superfluid components. Second, the thermodynamic potentials of the two components are equal. Third, the pressures are equal at the interface. Since the pressure is proportional to the thermodynamic potential in the present framework, it is sufficient to equate the thermodynamic potentials obtained for the normal and superfluid phases. The equality of both thermodynamic potentials (Eqs. (24) and (28)) gives

$$\begin{aligned} &\frac{1}{4}\mu_s^{5/2} m_+^{3/2}\left(1+X^2\right)^{1/4}\left[-4\sqrt{1+X^2}\,P_{3/2}\left(-\left(1+X^2\right)^{-1/2}\right)+\left\{\left(1+X^2\right)-5\right\}P_{1/2}\left(-\left(1+X^2\right)^{-1/2}\right)\right] \\ &=\frac{2}{3\pi}\left(\mu_s+U_s+h_s-U_{N_\uparrow}\right)^{5/2}+\frac{2\left(m_r\right)^{3/2}}{3\pi}\left(\mu_s+U_s-h_s-U_{N_\downarrow}\right)^{5/2} \end{aligned} \tag{29}$$

Eq. (29) and the self-consistency equations for the HF potentials in the normal component (Eqs. (21) and (22)) are solved simultaneously for $h_s$, $U_{N_\uparrow}$ and $U_{N_\downarrow}$ after $\mu_s$, $\Delta$, and $U_s$ have been obtained from the superfluid equations.

## C. Quasiparticle specific heat

The quasiparticle specific heat is evaluated separately in the superfluid and normal components and then added. First, the calculation of specific heat in the superfluid component is given. For the BCS regime, one can write the entropy of the fermionic system as [73-74]

$$s=-\sum_{\vec{k},\sigma=\alpha,\beta}\left(f\left(E_{k_\sigma}\right)\ln f\left(E_{k_\sigma}\right)+\left(1-f\left(E_{k_\sigma}\right)\right)\ln\left(1-f\left(E_{k_\sigma}\right)\right)\right) \tag{30}$$

where $f\left(E_{k_\sigma}\right)$ is the Fermi-Dirac distribution function with the excitation energy, $E_{k_\sigma}$. By using the Fermi-Dirac distribution function, the entropy becomes

$$s=\sum_{\vec{k}}\left(-\frac{E_{k_\alpha}}{T}\frac{e^{E_{k_\alpha}/T}}{1+e^{E_{k_\alpha}/T}}+\ln\left(1+e^{E_{k_\alpha}/T}\right)\right) \tag{31}$$

For calculating the specific heat, one can use

$$c_v=T\left(ds/dT\right) \text{ or } c_v=\left(d/dT\right)U=\left(d/dT\right)\sum_{\mathbf{k},\sigma=\alpha,\beta}E_{k_\sigma}f\left(E_{k_\sigma}\right) \tag{32}$$

The second expression in Eq. (32) is indeed the derivative of the internal energy, $\sum_{\mathbf{k},\sigma=\alpha,\beta}E_{k_\sigma}f\left(E_{k_\sigma}\right)$, with respect to temperature. Both formulas given in Eq. (32) yield the same result for the specific heat. If one ignores the temperature-dependence of the gap function existing in the $E_{k_\sigma}$, then, the specific heat for superfluid component, which identified by index S, is given by

$$\left(c_v\right)_S=\sum_{k}\frac{E_{k_\alpha}^2}{T^2}\left(1-f\left(E_{k_\alpha}\right)\right)f\left(E_{k_\alpha}\right) \tag{33}$$

It should be noted that the Andreev approximation will be used, i.e., $k^p=k_p=k^h=k_h$, in which quasiparticles (identified by index *p*) and quasiholes (identified by index *h*) in superfluid and normal parts are indicated by superscript and subscript indices, respectively. The summation can transform into the integration, i.e., $\sum_{\mathbf{k}} \to \int k^{\|} dk^{\|} dk^{p} d\varphi$. Then, one can consider the transformation of cylindrical coordinate $\left(k^{p}, k^{\|}\right)$ to $\left(k_{p}, E\right)$, by using the relation between $k^{\|}$ and $E$ via the formula $k^{\|^{2}} = k_F^2 - k_p^2 + 2m_{\uparrow} E_{\alpha}$ (for $\alpha$ channel). Also, the energy $E$ can be related to energy $\varepsilon_{\alpha}$, via the formula $E_{\alpha} = \left(1/(1+m_r)\right)\left[2\varepsilon_{\alpha} - (1-m_r)\mu_s - h_s(1+m_r)\right]$. It should be mentioned that the minimum value of $\varepsilon_{\alpha}$ is $\varepsilon_{\alpha}^{0} = (1/2)\left((1+m_r)h_s\right) + \left((1-m_r)\mu_s\right)$.Also, the condition $\varepsilon_{\alpha} > \sqrt{m_r}\Delta$ should be established to exist quasiparticles. Finally, $(c_v)_S$ is given by

$$(c_v)_S = \frac{m_{\uparrow} k_F}{2\pi^2 (1+m_r) T^2} \int_{\varepsilon_{\alpha}^{0}}^{d} \frac{(a\varepsilon_{\alpha} + b)^2 e^{(a\varepsilon_{\alpha}+b)/T}}{\left(1+e^{(a\varepsilon_{\alpha}+b)/T}\right)^2} d\varepsilon_{\alpha} \tag{34}$$

where

$$a = \frac{2}{1+m_r}\ ,\ b = \frac{-\mu_s(1-m_r)}{1+m_r}\ ,\ d = \frac{1}{2}\left[(h_s + e_{\alpha})(1+m_r) + \mu_s(1-m_r)\right]$$
$$e_{\alpha} = -h_s + \frac{m_+}{m_-} + \sqrt{\left(\frac{k_F^2}{2m_+} - \mu_s\right)^2 + \Delta^2} \tag{35}$$

Furthermore, as a consequence of taking into account the temperature dependence of the energy gap and transforming $\Delta \to \Delta - \sqrt{2\pi T \Delta}\left(1-(T/8\Delta)\right)e^{-\Delta/T}$, Eq. (33) needs to have an additional term $-(1/T)\sum_{k}\left(E_{k_\alpha} dE_{k_\alpha}/dT\right)\left(1-f\left(E_{k_\alpha}\right)\right) f\left(E_{k_\alpha}\right)$, where $dE_{k_\alpha}/dT$ can be replaced by $(\Delta d\Delta/dT)\Big/\sqrt{\left(\left(k^2/2m_+\right)-\mu_s\right)^2+\Delta^2}$. However, the numerical calculation shows that the temperature dependence of the energy gap does not affect the specific heat at the temperatures under study. It should be noted that the energy gap does not depend on mass ratio and does not change for different mixtures under the same conditions.

Now, the calculation of the specific heat in normal component, which identified by the index N, is given. The quasiparticle energy is given by

$$E_{N_\uparrow} = \frac{k^2}{2m_\uparrow} - \mu_\uparrow + U_{N_\uparrow} = \frac{k^2}{2m_\uparrow} - \mu_s - h_s - U_s + U_{N_\uparrow} \tag{36}$$

$$E_{N_\downarrow} = \frac{k^2}{2m_\downarrow} - \mu_\downarrow + U_{N_\downarrow} = \frac{k^2}{2m_\downarrow} - \mu_s + h_s - U_s + U_{N_\downarrow} \tag{37}$$

Using Eqs. (36) and (37) together with Eq. (33) for the normal component, by replacing $E_{k_\sigma}$ by $E_{N_i}$ ($i =\uparrow$ and $\downarrow$), $(c_v)_N$ becomes

$$
\begin{aligned}
(c_v)_N &= \frac{\sqrt{2}\left(m_\uparrow\right)^{3/2}}{2\pi^2 T^2}\int_0^{\mu_s+h_s+U_s-U_{N\uparrow}}\left(E_{N_\uparrow}+\mu_s+h_s+U_s-U_{N_\uparrow}\right)^{1/2}\frac{e^{E_{N\uparrow}/T}}{\left(1+e^{E_{N\uparrow}/T}\right)^2}dE_{N_\uparrow} \\
&+\frac{\sqrt{2}\left(m_r\right)^{3/2}}{2\pi^2 T^2}\int_0^{\mu_s-h_s+U_s-U_{N\downarrow}}\left(E_{N_\downarrow}+\mu_s-h_s+U_s-U_{N_\downarrow}\right)^{1/2}\frac{e^{E_{N\downarrow}/T}}{\left(1+e^{E_{N\downarrow}/T}\right)^2}dE_{N_\downarrow}
\end{aligned} \tag{38}
$$

Eq. (38) is used to calculate numerically the specific heat of the normal component. To calculate the specific heat, the values of $h_s, \mu_s$ and $\Delta$ are required, which are obtained numerically from the self-consistent equations described above. Now, by using the relevant physical parameters, i.e., $\mu_s$, $\Delta$ and $h_s$, which are obtained numerically, the quasiparticle specific heat (Eqs. (34) and (38)) can be plotted and interpreted.

## III. Numerical calculations and results

### A-Phase separation parameters

The numerical analysis begins with the quantities that specify the normal-separated state. Fig. 2 shows the mass-ratio dependence of the required imbalance chemical potential and, for the $^{163}$Dy-$^{40}$K mixture, the dependence of $h_s$, $\mu_s$ and $\Delta$ on the interaction strength. These quantities are obtained from the same self-consistent together with the phase-separation condition.

Fig. 2 (a) shows that the dimensionless imbalance chemical potential (in units of the Fermi energy, $h_s / E_F$) required for normal-superfluid phase separation increases with the mass ratio, $m_r$, at a fixed interaction strength. The values obtained in Fig. 2 are valid for $h_s < \Delta$, because otherwise other phases, such as the Sarma phase, occur. With a larger mass ratio, polarization has a stronger effect on the pressure in the normal region, while kinematic mismatch simultaneously weakens the pairing contribution to the superfluid. The phase-separation condition (equal pressures across the N–SF interface) therefore requires a larger imbalance chemical potential so that the normal component can maintain the pressure balance.

For the $^{163}$Dy-$^{40}$K mixture ($m_r = 4.08$), Fig. 2 (b) and its inset show that with increasing $|1/k_F a|$, both imbalance and average chemical potentials increase, whereas Fig. 2(c) shows that the energy gap decreases. These accompanying changes in $h_s$, $\mu_s$, and $\Delta$ modify the energy spectrum and, consequently, the superfluid density of states, as discussed later.

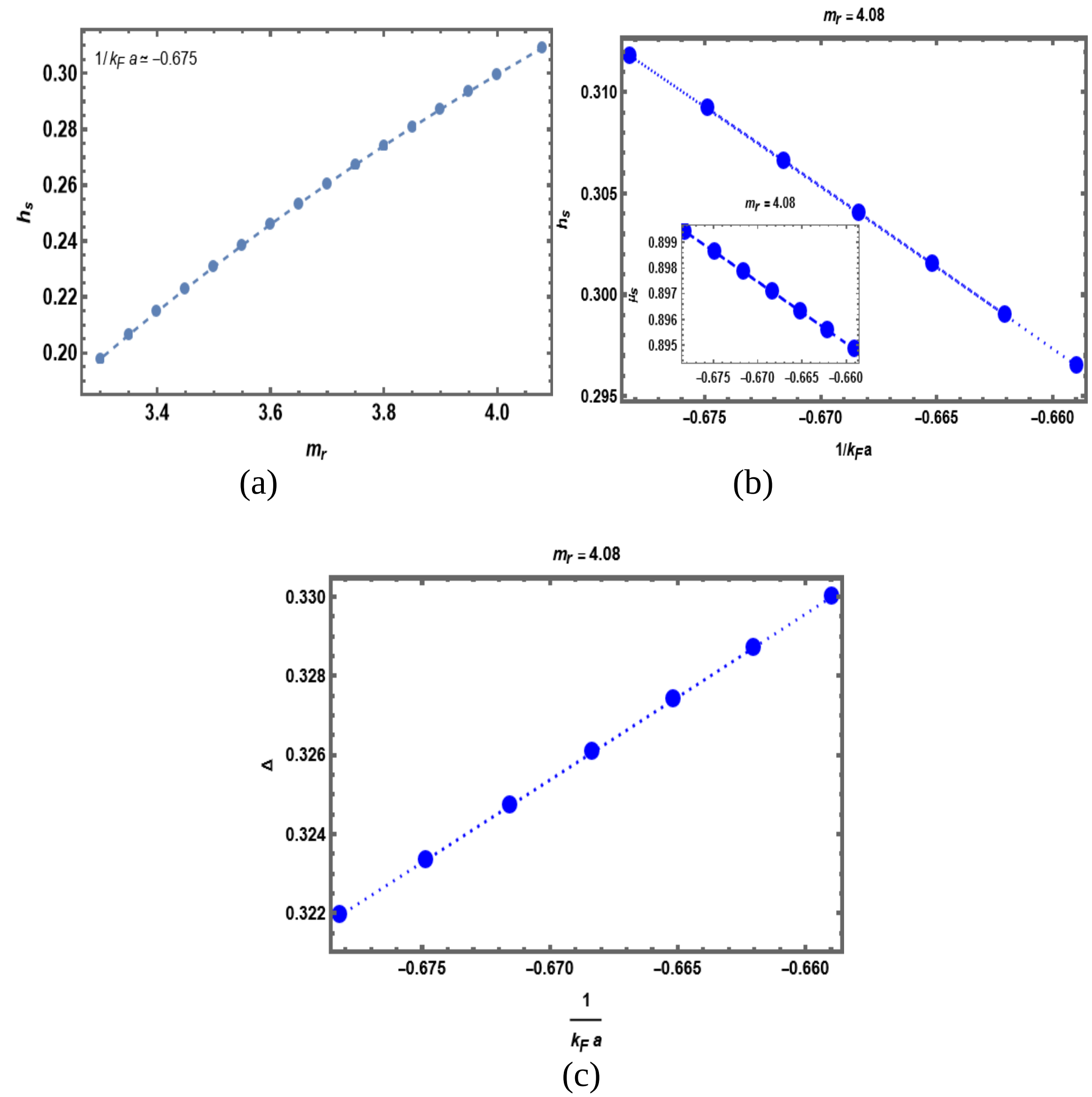

**Fig. 2.** (Color online) Phase-separation parameters. (a) The dimensionless imbalance chemical potential (in units of the Fermi energy, $h_s / E_F$ ) versus the mass ratio, $m_r$ , at a fixed interaction strength, $1/k_F a \simeq -0.675$. (b) The dimensionless imbalance chemical potential (in units of the Fermi energy, $h_s / E_F$ ) versus the interaction strength, $1/k_F a$ at the mass ratio $m_r = 4.08$ corresponding to the $^{163}$Dy-$^{40}$K mixture. Inset: The dimensionless average chemical potential (in units of the Fermi energy, $\mu_s / E_F$ ) versus the interaction strength, $1/k_F a$ , at the mass ratio $m_r = 4.08$. (c) The dimensionless energy gap (in units of the Fermi energy, $\Delta / E_F$ ) versus the interaction strength, $1/k_F a$ , at the mass ratio $m_r = 4.08$.

It is worth noting that the zero-temperature limit has been used to calculate the relevant parameters, $\Delta$, $\mu_s$, and $h_s$, as well as the grand-canonical thermodynamic potentials of the superfluid and normal components. The validity of this limit at finite temperature is discussed below.

As long as the conditions $T << \Delta$ and $T << \mu_\sigma - U_{N_\sigma}$ hold, the zero-temperature equations provide a reliable approximation at finite temperatures with a minor error. For example, the finite-temperature gap equation was presented in Eq. (12). Under the condition, $T << \Delta$, the effect of the Fermi–Dirac distribution functions in Eq. (12) is small, so the equation can be reduced by its zero-temperature form .In the normal component, for example, the particle density at finite temperature for each spin is given by Eq. (17). When $T << \mu_\sigma - U_{N_\sigma}$, only a negligible number of particles occupy states above the level $\mu_\sigma - U_{N_\sigma}$. The Fermi-Dirac function in Eq. (17) then approaches its zero-temperature step-function form, and the particle density equals its zero-temperature value given by Eq. (20). Throughout the system considered, these conditions hold; therefore the zero-temperature equations are used. Only the conditions $T << \Delta$ and $T << \mu_\sigma - U_{N_\sigma}$ should be checked. For this purpose, it is stated that the maximum temperature considered in this paper is $0.02T_F$, while the typical energy gap is roughly ten times the maximum temperature (Fig. 2(c)). hence $T << \Delta$. Moreover, using $\mu_\uparrow - U_{N_\uparrow} \equiv \mu_s + h_s + U_s - U_{N_\uparrow}$, indicates that $T << \mu_\uparrow - U_{N_\uparrow}$ is satisfied (the same argument holds for $T << \mu_\downarrow - U_{N_\downarrow}$) .

It is important to know the critical temperature, namely the transition from the phase-separated superfluid–normal state to a fully normal state, since the chosen parameters lie within the phase-separated regime. However, determining this temperature requires solving the finite-temperature equations self-consistently, which is not trivial, and therefore a qualitative approach is used. For each mass ratio and interaction strength, the critical temperature can be determined for every value of the polarization[38-39,75-76]. By working at temperatures below $T/T_F \cong 0.02$, the system remains well below the critical temperature for all parameter sets, and the normal–superfluid phase-separated regime is maintained[77] for all mass ratios and all interaction strengths considered here.

## B-**Superfluid density of states**

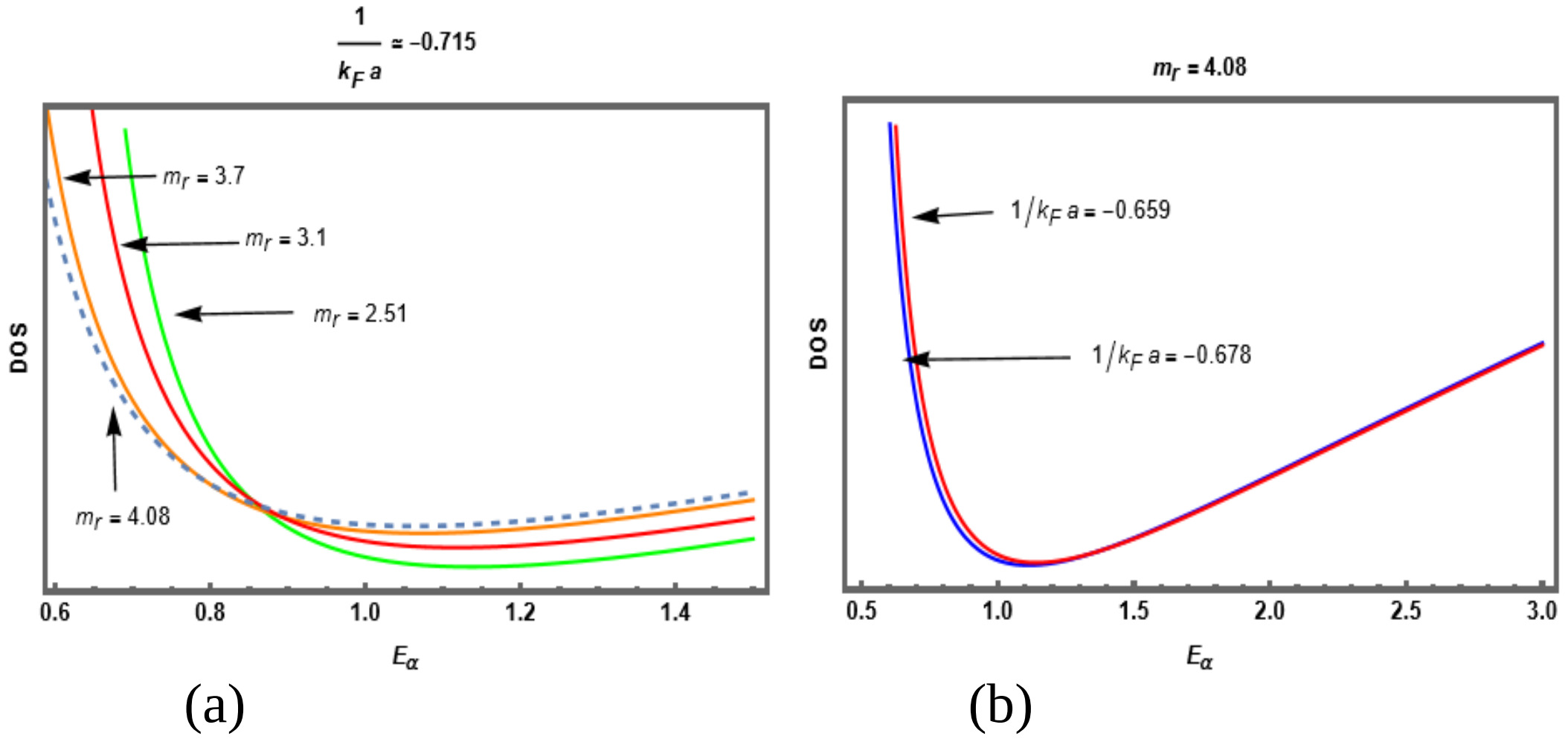

**Fig. 3.** (Color online) (a) The superfluid density of states (DOS) in arbitrary units versus the excitation energy, at a fixed interaction strength, $1/k_F a \simeq -0.715$, and at different mass ratios, $m_r = 2.51, 3.1, 3.7$ and $4.08$. (b) The superfluid density of states (DOS) in arbitrary units versus the excitation energy, at a mass ratio, $m_r = 4.08$, at different interaction strengths, $1/k_F a \simeq -0.659$, and $-0.678$.

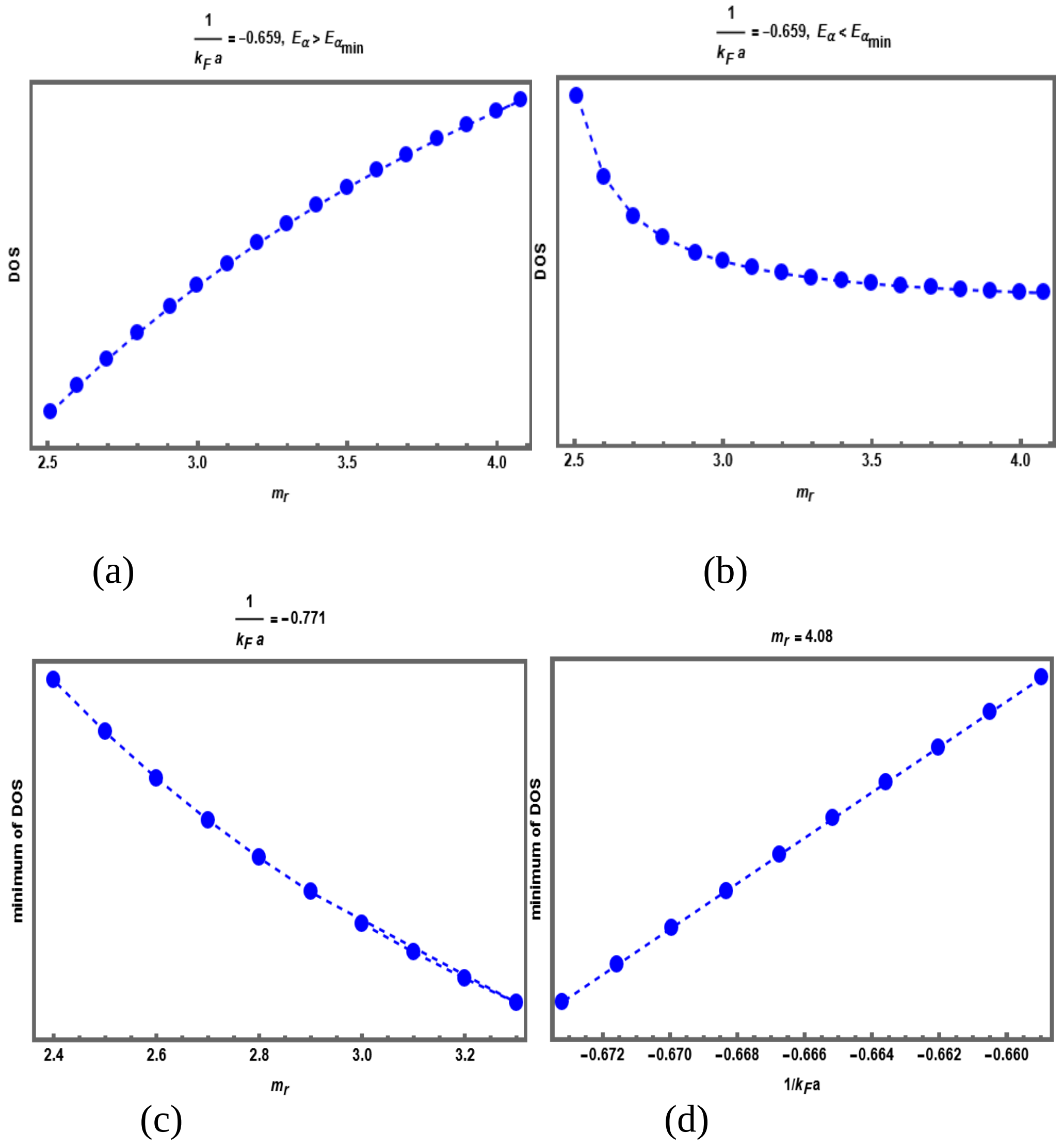

**Fig. 4.** (Color online) Dependence of the superfluid density of states (DOS) in arbitrary units on mass ratio, $m_r$ and interaction strength, $1/k_F a$ . (a) The superfluid DOS (in arbitrary units) versus $m_r$ at $1/k_F a \simeq -0.659$ for a selected excitation energy $E_\alpha > E_{\alpha_{\min}}$ (here $E_{\alpha_{\min}}$ denotes the excitation energy at which the DOS reaches its minimum) . (b) The superfluid DOS (in arbitrary units) versus $m_r$ at $1/k_F a \simeq -0.659$ for a selected excitation energy $E_\alpha < E_{\alpha_{\min}}$ . (c) Minimum DOS (in arbitrary units) versus $m_r$ at $1/k_F a \simeq -0.771$. (d) Minimum DOS (in arbitrary units) versus $1/k_F a$ at $m_r = 4.08$.

Fig. 3(a) shows the superfluid DOS in arbitrary units versus the excitation energy at a fixed interaction strength, $1/k_F a \simeq -0.715$, and for different mass ratios, $m_r = 2.51, 3.1, 3.7$ and $4.08$. For the $^{163}$Dy-$^{40}$K mixture, with $m_r = 4.08$, Fig. 3 (b) shows superfluid DOS (in arbitrary units) versus the excitation energy at two different interaction strengths, $1/k_F a \simeq -0.659$, and $-0.678$. Fig. 4 (a) and 4(b) show the superfluid DOS (in arbitrary units) versus mass ratio, $m_r$, at $1/k_F a \simeq -0.659$ for selected excitation energies above and below $E_{\alpha_{\min}}$, respectively, where $E_{\alpha_{\min}}$ denotes the excitation energy at which the superfluid DOS reaches its minimum. The superfluid DOS increases with $m_r$ in Fig. 4(a), whereas it decreases with $m_r$ in Fig. 4(b). The minimum of density of states versus mass ratio, $m_r$, at $1/k_F a \simeq -0.771$ is shown in Fig. 4(c), and it decreases with increasing $m_r$. For the $^{163}$Dy-$^{40}$K mixture, the minimum of the superfluid DOS versus the interaction strength, $1/k_F a$, is shown in Fig. 4(d). The superfluid DOS, which is not plotted to avoid repetition, also shows the same behavior as the minimum of the superfluid DOS as a function of $1/k_F a$. Thus, the superfluid DOS decreases with increasing absolute value of the interaction strength**.** On the BCS side, this increase corresponds to a decrease in the absolute value of the scattering length and hence to weaker pairing.

To understand the results shown in Fig. 2, the changes in the superfluid DOS can also be considered. The decrease in the superfluid DOS with increasing absolute value of the interaction strength accompanies the reduction of the pairing gap. This drop in the gap leads to a lower condensation energy, which in turn reduces the thermodynamic pressure of the superfluid phase (which is proportional to $\Omega_s$). In the phase-separated state, mechanical equilibrium requires that the superfluid and normal phases have the same pressure. To compensate for the drop in superfluid pressure, the system increases the average chemical potential, which primarily controls the total particle number density. A higher $\mu_s$ leads to a higher particle number density in the superfluid component and restores the superfluid pressure. However, increasing $\mu_s$ also raises the pressure of the normal phase. To keep the exact balance between the two phases, the system must also adjust normal pressure. This is achieved by adjusting the imbalance chemical potential, which directly controls the relative population of the two spin components. In the present case, the polarization of the normal component, defined by $\left(n_\uparrow^n - n_\downarrow^n\right)/\left(n_\uparrow^n + n_\downarrow^n\right)$, is negative. Consequently, an increasing imbalance chemical potential, acts to reduce the population of these majority down-spins. This reduction in the down-spin population lowers the overall pressure of the normal phase. Therefore, the simultaneous increase in $h_s$ is a necessary mechanism to adjust the normal phase pressure downwards, ensuring the system remains in mechanical equilibrium.

## C-**Quasiparticle specific heat**

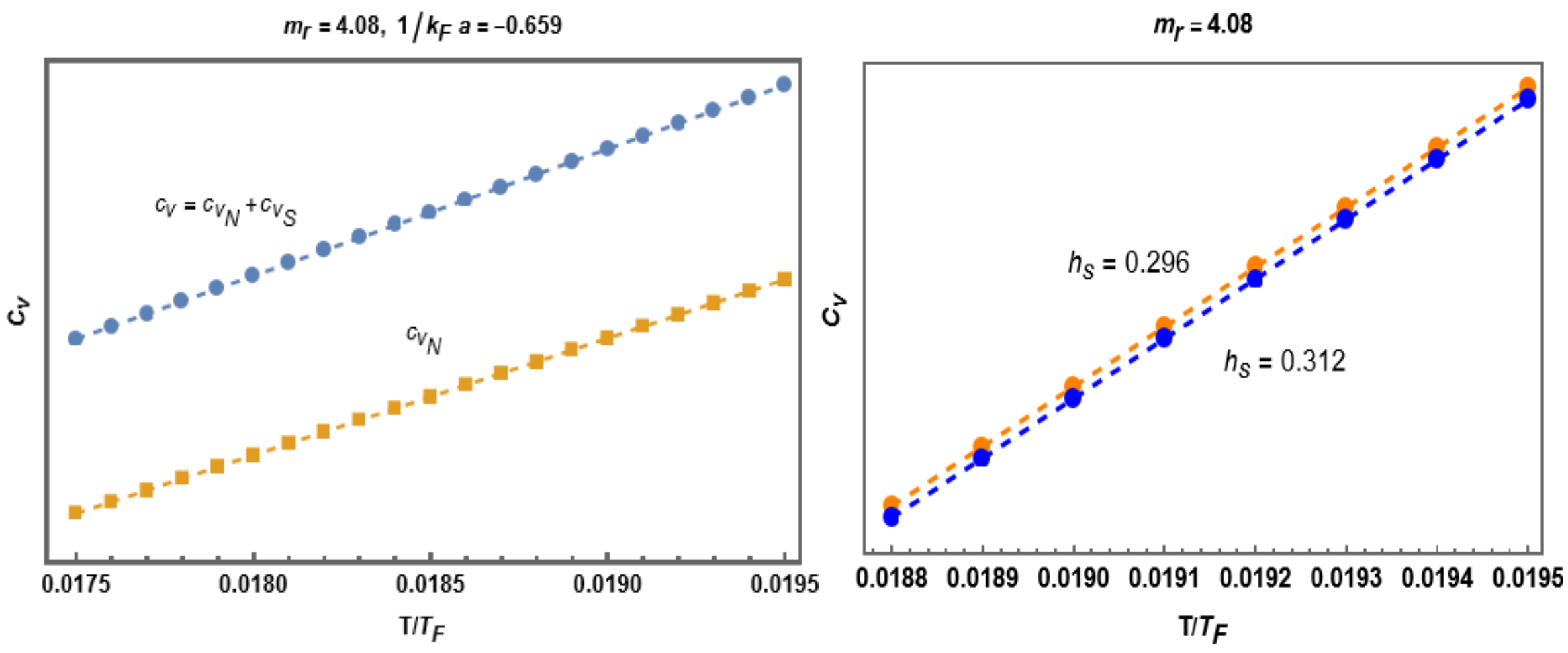

(a) (b)

**Fig. 5**. (Color online)Temperature dependence of the quasiparticle specific heat. (a) Total specific heat, $c_v$ ($\equiv c_{v_N} + c_{v_S}$ where $c_{v_S}$ and $c_{v_N}$ are superfluid and normal specific heat, respectively) and normal specific heat, $c_{v_N}$, (in arbitrary units) versus $T/T_F$ at a fixed mass ratio, $m_r = 4.08$, and at a fixed interaction strength, $1/k_F a \simeq -0.659$. (b)Total $c_v$ (in arbitrary units) versus $T/T_F$ at a fixed mass ratio, $m_r = 4.08$, and at two different dimensionless imbalance chemical potentials (in units of the Fermi energy, $h_s / E_F$), $h_s = 0.296$ and $0.312$ corresponding to $1/k_F a = -0.659$ and $-0.678$, respectively.

Now the specific heat is investigated on different relevant parameters. Fig. 5(a) shows total specific heat, $c_v$ ($\equiv c_{v_N} + c_{v_S}$ where $c_{v_S}$ and $c_{v_N}$ are superfluid and normal specific heat, respectively) and normal specific heat, $c_{v_N}$, (in arbitrary units) as a function of $T/T_F$ at a fixed mass ratio, $m_r = 4.08$, and at a fixed interaction strength, $1/k_F a \simeq -0.659$. The normal component gives the dominant contribution to the total specific heat, while the superfluid component provides a smaller contribution. At a fixed temperature and a fixed mass ratio, $m_r = 4.08$, for the lower value of $h_s$, total specific heat, $c_v$ is higher than that of the higher value of $h_s$ ( Fig. 5(b)). For the two cases in Fig. 5(b), the two values of $h_s$ are those obtained from the phase-separation calculation at the corresponding interaction strengths; they are not varied independently in this comparison. It should be mentioned that the change of $h_s$ can be created by the change in mass ratios or interaction strengths. Therefore, because of considering a constant mass ratio in Fig. 5(b), the change of $h_s$ is due to the change of interaction strength; in the fixed other parameters, $h_s = 0.296\,(0.312)$ is accompanied with $1/k_F a = -0.659$ (-0.678). A plot of $c_v$ with respect to $h_s$ (and $1/k_F a$) at a fixed value of mass ratio, $m_r = 4.08$ and at a fixed temperature, $T/T_F = 0.019$, is shown in Fig. 6(a) (and the inset of Fig. 6 (a)). An increase in $h_s$ or $|1/k_F a|$ leads to a decrease in $c_v$.

The following points provide the basis for the explanation. Increasing the absolute value of the interaction strength makes pair formation more difficult and is accompanied by a decrease in the superfluid DOS, which reduces the quasiparticle contribution to the heat capacity. As a consequence, the superfluid component provides only a small contribution to the heat capacity. Calculations confirm that, compared with the normal component, its contribution is negligible; the heat capacity is therefore mainly governed by the normal phase. In the normal component, the spin-down particles not only form the majority of the population but also have a larger mass. The heavier spin-down component has a larger density of available states near the Fermi surface. As a result, quasiparticles close to the Fermi surface of the spin-down species act as the primary source of thermal excitations. Furthermore, as previously discussed, when the magnitude of the interaction strength increases, the superfluid pressure decreases. To maintain mechanical equilibrium between the phases and pressure equality, the system must increase both the average chemical potential and the imbalance chemical potential. Because the increase in $h_s$ is dominant, the net spin-down chemical potential ($\mu_\downarrow = \mu_s + U_s - h_s$) decreases. According to Eq. (20), this reduction leads to a decrease in the number density of the spin-down particles (see also Ref. [77]). Consequently, the population of these crucial thermal quasiparticles is reduced, which directly leads to a smaller total heat capacity.

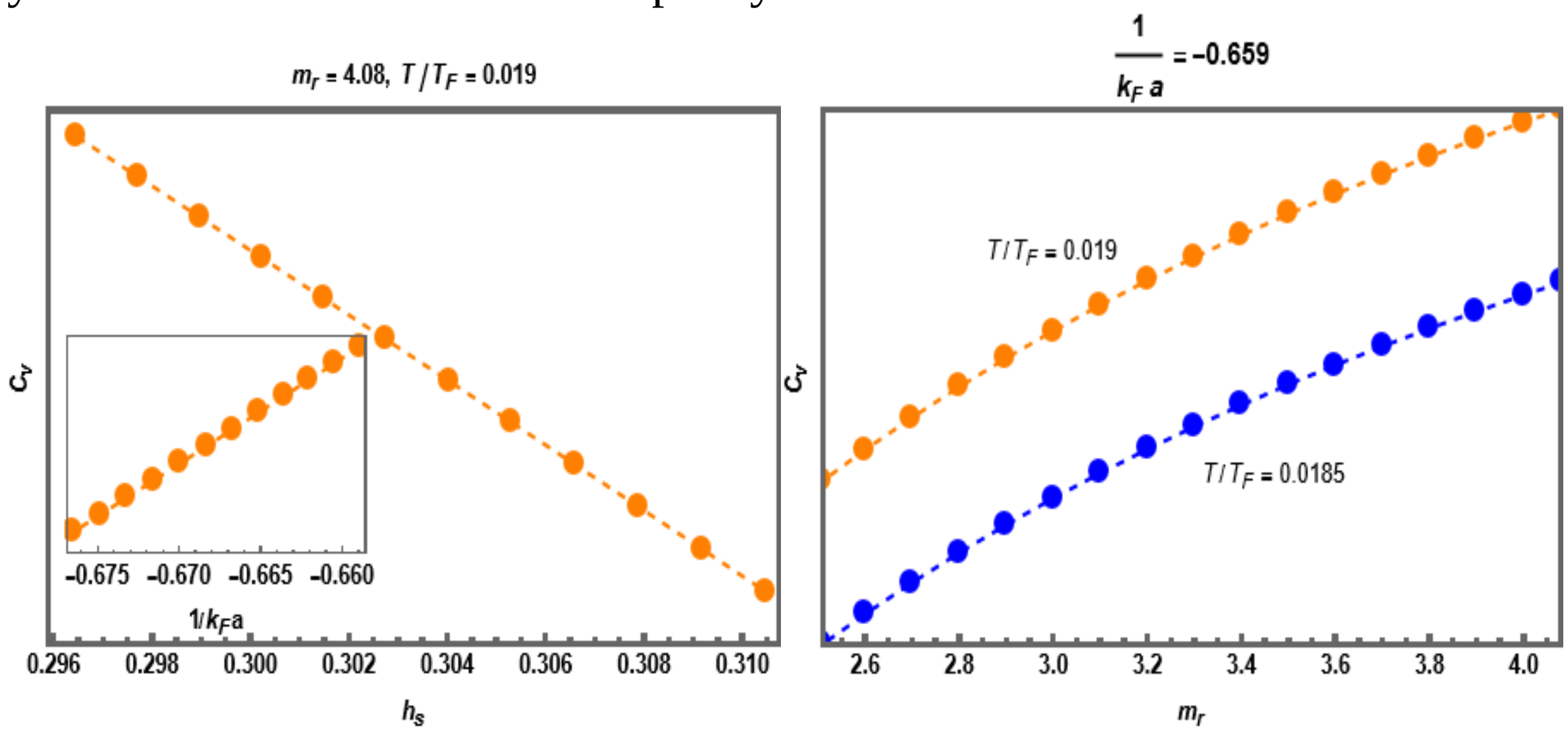

(a) (b)

**Fig. 6.** (Color online) Dependence of the specific heat (in arbitrary units) on the dimensionless imbalance chemical potential (in units of the Fermi energy, $h_s / E_F$), interaction strength, and mass ratio. (a) $c_v$ versus $h_s$ for $m_r = 4.08$ at a fixed

temperature $T/T_F = 0.019$. Inset: the corresponding $c_v$ versus $1/k_F a$ . (b) $c_v$ versus mass ratio, $m_r$, at two different temperatures $T/T_F = 0.0185$ and $0.019$ for $1/k_F a = -0.659$.

Fig.6(b) illustrates how $m_r$ affects $c_v$. Fig. 6 (b) shows $c_v$ increases with $m_r$ at two different temperatures $T/T_F = 0.0185$ and 0.019 for $1/k_F a = -0.659$; the effect of $m_r$ on the enhancement of $c_v$ is more evident when the temperature is raised. This means that Fermi-Fermi mixtures with a higher mass ratio have higher specific heat. The reason is as follows. When the mass ratio increases, the heavier spin-down component provides a larger density of available states near its own Fermi surface. Since the quasiparticle specific heat is controlled by excitations near the Fermi surface, this increase in the density of states due to the larger mass can enhance the quasiparticle contribution to the specific heat, especially in the phase-separated regime.

## IV. Conclusions

The thermodynamic properties of mass-imbalanced spin-polarized Fermi–Fermi mixtures have been studied in the normal–superfluid phase-separated regime, with particular emphasis on the quasiparticle specific heat. The $^{163}$Dy–$^{40}$K mixture, with $m_r = 4.08$, is used as the principal example. The analysis is restricted to the BCS side of the BCS–BEC crossover and to $h_s < \Delta$, so that Sarma-type polarized-superfluid solutions are excluded. Normal–superfluid phase separation is determined from chemical and mechanical equilibrium rather than by imposing a Clogston relation for the unequal-mass mixtures. The chemical potentials of each spin species are equal in the normal and superfluid phases, and equality of the grand-canonical thermodynamic potentials of the two phases determines the phase-separation condition and the corresponding imbalance chemical potential. The resulting phase diagram shows that, at fixed interaction strength, the required $h_s$ increases as $m_r$ increases. Thus, mass imbalance directly shifts the phase-separation condition even before the thermal response is considered. For the $^{163}$Dy–$^{40}$K mixture, increasing the magnitude of the dimensionless interaction strength over the BCS-side interval considered here is accompanied by increases in the average chemical potential $\mu_s$ and the imbalance chemical potential determined by the phase-separation condition, $h_s$, while the pairing gap $\Delta$ decreases. For each interaction strength, $\Delta$, $\mu_s$, and $U_s$ are obtained from the superfluid equations, and the corresponding $h_s$ and normal-state HF potentials follow from the phase-separation condition and normal-state self-consistency equations. The decrease in the pairing gap changes the superfluid thermodynamic potential and pressure. Mechanical equilibrium with the normal component therefore requires corresponding changes in the average and imbalance chemical potentials. The superfluid density of states was examined as a function of excitation energy, interaction strength, and mass ratio. For the $^{163}$Dy–$^{40}$K mixture, both the DOS and its minimum decrease with increasing magnitude of the interaction strength. The minimum DOS also decreases with increasing mass ratio. The changes in the DOS, together with the pressures of the normal and superfluid components, are used to explain the corresponding changes in the relevant thermodynamic parameters and to interpret the behavior of the specific heat. The quantities used in the calculated specific heat were obtained from the zero-temperature forms of the gap, number, Hartree–Fock, and phase-separation equations. This approximation is justified over the numerical temperature range by the conditions $T \ll \Delta$ and $T \ll \mu_\sigma - U_{N\sigma}$. At low temperatures, the normal component provides the dominant contribution to the total specific heat. For a fixed mass ratio, the calculated specific heat decreases with increasing the corresponding $h_s$ or the absolute value of $1/k_F a$. In the phase-separated states considered here, the spin-down particles form the majority population and are also the heavier species, so they provide the principal low-energy thermal excitations. Mechanical equilibrium requires corresponding changes of the chemical potentials. Over the parameter range considered, the effective spin-down chemical potential $\mu_\downarrow = \mu_s + U_s - h_s$ decreases, reducing the spin-down number density and therefore the dominant normal-state contribution to the specific heat. Increasing $m_r$ increases the total quasiparticle specific heat because the heavier spin-down component supplies a larger density of available states near its Fermi surface. This enhancement is more pronounced at the higher temperatures within the low-temperature range studied. The phase-diagram results, the quasiparticle DOS, and the specific-heat results show that the thermal response provides a characteristic signature of mass imbalance and normal–superfluid phase separation. The conclusions are restricted to the mean-field, BCS-side, low-temperature regime considered here.

**Acknowledgment**
I am grateful to Prof. A. J. Leggett for several useful discussions.

**Data Availability Statement**
All data generated or analyzed during this study are included in this published article.

## Appendix A:

To obtain the effective Hamiltonian in terms of the Bogoliubov operators, $\gamma^\dagger$ and $\gamma$, the derivation can be carried out through the following steps. Substitution of the inverse Bogoliubov transformation into the momentum-space Hamiltonian gives

$$\sum_k \gamma_{k\alpha}^{\dagger}\gamma_{k\alpha}\left(u_k^2\left(\frac{k^2}{2m_\uparrow}-\mu_\uparrow+U_\uparrow\right)-v_k^2\left(\frac{k^2}{2m_\downarrow}-\mu_\downarrow+U_\downarrow\right)-2\Delta u_k v_k\right)$$
$$+\sum_k \gamma_{k\beta}^{\dagger}\gamma_{k\beta}\left(-v_k^2\left(\frac{k^2}{2m_\uparrow}-\mu_\uparrow+U_\uparrow\right)+u_k^2\left(\frac{k^2}{2m_\downarrow}-\mu_\downarrow+U_\downarrow\right)-2\Delta u_k v_k\right)$$
$$+\sum_k \gamma_{k\alpha}^{\dagger}\gamma_{k\beta}^{\dagger}\left(u_k v_k\left(\frac{k^2}{2m_\uparrow}-\mu_\uparrow+U_\uparrow\right)+u_k v_k\left(\frac{k^2}{2m_\downarrow}-\mu_\downarrow+U_\downarrow\right)+\left(u_k^2-v_k^2\right)\Delta\right) \quad \text{(A1)}$$
$$+\sum_k \gamma_{k\beta}\gamma_{k\alpha}\left(\left(\frac{k^2}{2m_\uparrow}-\mu_\uparrow+U_\uparrow\right)u_k v_k+\left(\frac{k^2}{2m_\downarrow}-\mu_\downarrow+U_\downarrow\right)u_k v_k+\left(u_k^2-v_k^2\right)\Delta\right)$$
$$+const$$

For a diagonalized Hamiltonian, it is clear that the coefficients of $\gamma_{k\alpha}^{\dagger}\gamma_{k\beta}^{\dagger}$ and $\gamma_{k\alpha}\gamma_{k\beta}$ should be set to zero, i.e.,

$$u_k v_k\left(\frac{k^2}{2m_\uparrow}+\frac{k^2}{2m_\downarrow}-2\mu_s-2U_s+U_\uparrow+U_\downarrow\right)+\left(u_k^2-v_k^2\right)\Delta=0 \quad \text{(A2)}$$

Using the definition of $m_+\left(\equiv 2\left(m_\uparrow^{-1}+m_\downarrow^{-1}\right)^{-1}\right)$ and supposing that $U_s=U_\uparrow=U_\downarrow$, Eq. (A2) becomes

$$2u_k v_k\left(\frac{k^2}{2m_+}-\mu_s\right)=-\left(u_k^2-v_k^2\right)\Delta \quad \text{(A3)}$$

Squaring the equation and using $u_k^2+v_k^2=1$ gives

$$4u_k^2v_k^2\left(\frac{k^2}{2m_+}-\mu_s\right)^2=\Delta^2\left(\left(u_k^2+v_k^2\right)^2-4u_k^2v_k^2\right)=\Delta^2\left(1-4u_k^2v_k^2\right) \quad \text{(A4)}$$

and therefore

$$u_k v_k=\frac{\Delta}{2\sqrt{\left(\frac{k^2}{2m_+}-\mu_s\right)^2+\Delta^2}} \quad \text{(A5)}$$

The combination of Eq. (A5) and the normalization condition yields

$$v_k^2=\frac{1}{2}\left(1-\frac{\left(\frac{k^2}{2m_+}-\mu_s\right)}{\sqrt{\left(\frac{k^2}{2m_+}-\mu_s\right)^2+\Delta^2}}\right) \quad \text{and} \quad u_k^2=\frac{1}{2}\left(1+\frac{\left(\frac{k^2}{2m_+}-\mu_s\right)}{\sqrt{\left(\frac{k^2}{2m_+}-\mu_s\right)^2+\Delta^2}}\right) \quad \text{(A6)}$$

where the average chemical potential, $\mu_s$, is defined as $\mu_s=\left((\mu_\uparrow+\mu_\downarrow)/2\right)-U_s$. With $u_k$ and $v_k$, the off-diagonal coefficients in Eq. (A1) vanish, and the coefficients of $\gamma^{\dagger}\gamma$ are the energies of the $\alpha$ and $\beta$ channels, $E_{k_\alpha}$ and $E_{k_\beta}$, respectively.

$$\left(u_k^2\left(\frac{k^2}{2m_\uparrow}-\mu_\uparrow+U_\uparrow\right)-v_k^2\left(\frac{k^2}{2m_\downarrow}-\mu_\downarrow+U_\downarrow\right)-2\Delta u_k v_k\right)$$
$$=-h_s+\frac{k^2}{2m_-}+\sqrt{\left(\frac{k^2}{2m_+}-\mu_s\right)^2+\Delta^2}\equiv E_{k_\alpha} \tag{A7}$$

$$\left(-v_k^2\left(\frac{k^2}{2m_\uparrow}-\mu_\uparrow+U_\uparrow\right)+u_k^2\left(\frac{k^2}{2m_\downarrow}-\mu_\downarrow+U_\downarrow\right)-2\Delta u_k v_k\right)$$
$$=h_s-\frac{k^2}{2m_-}+\sqrt{\left(\frac{k^2}{2m_+}-\mu_s\right)^2+\Delta^2}\equiv E_{k_\beta} \tag{A8}$$

Also, using Eqs. (A5) and (A6), the c-number term in Eq. (A1) becomes

$$\sum_k \left(\frac{k^2}{2m_\uparrow}-\mu_\uparrow+U_\uparrow\right)v_k^2+\left(\frac{k^2}{2m_\downarrow}-\mu_\downarrow+U_\downarrow\right)v_k^2-2u_k v_k\Delta=$$
$$=\sum_k \left(\frac{k^2}{2m_+}-\mu_s\right)-\sqrt{\left(\frac{k^2}{2m_+}-\mu_s\right)^2+\Delta^2} \tag{A9}$$

The last term in Eq. (6) follows from the mean-field approximation . for two operators A and B

$$AB\cong\langle A\rangle B+A\langle B\rangle-\langle A\rangle\langle B\rangle \tag{A10}$$

where $\langle ...\rangle$ denotes the quantum-mechanical expectation value, for pairing interaction, the resulting constant term is

$$-g\sum_{\vec{k},\vec{k}'}\langle a_{\vec{k}\uparrow}^\dagger a_{-\vec{k}\downarrow}^\dagger\rangle\langle a_{-\vec{k}'\downarrow}a_{\vec{k}'\uparrow}\rangle \tag{A11}$$

Using the definition of the gap, i.e., $\Delta=-g\sum_{\vec{k}}\langle a_{-k\downarrow}a_{k\uparrow}\rangle$ and taking $\Delta$ to be real, this term becomes $-\Delta^2/g$ .

## Appendix B. Zero-temperature gap and number equations

Using $g=-4\pi|a|/m_+$ and the following equation[71,73]

$$\int_0^\infty\frac{z^\delta dz}{\sqrt{(z-1)^2+X^2}}=\frac{-\pi}{\sin\pi\delta}\left(1+X^2\right)^{\delta/2}P_\delta\left(-\frac{1}{\sqrt{1+X^2}}\right) \tag{B1}$$

where $P_\delta$ denotes the polynomial Legendre of degree $\delta$ , Eq. (13) becomes

$$\frac{1}{m_+a^2}=2\mu_s\sqrt{1+X^2}P_{1/2}^2\left(\frac{-1}{\sqrt{1+X^2}}\right) \tag{B2}$$

where $X = \mu_s/\Delta$. At finite temperature, the number equations for the two spin populations in the unpolarized superfluid are

$$n_{s_\uparrow} = \int_0^\infty \frac{d^3k}{(2\pi)^3}\left(u_k^2 f\left(E_\alpha\right) + v_k^2\left(1 - f\left(E_\beta\right)\right)\right) \tag{B3}$$

$$n_{s_\downarrow} = \int_0^\infty \frac{d^3k}{(2\pi)^3}\left(u_k^2 f\left(E_\beta\right) + v_k^2\left(1 - f\left(E_\alpha\right)\right)\right) \tag{B4}$$

After substituting $v_k^2$ and $u_k^2$, given by Eq. (A6), into Eqs. (B3) and (B4), total number equation for superfluid component becomes

$$n_s = n_{s_\uparrow} + n_{s_\downarrow} = \int_0^\infty \frac{d^3k}{(2\pi)^3}\left(1 - \frac{\left(k^2/2m_+ - \mu_s\right)\left(1 - f\left(E_\alpha\right) - f\left(E_\beta\right)\right)}{\sqrt{\left(k^2/2m_+ - \mu_s\right)^2 + \Delta^2}}\right) \tag{B5}$$

and in the zero-temperature limit

$$n_s\left(T=0\right) = n_{s_\uparrow}\left(T=0\right) + n_{s_\downarrow}\left(T=0\right) = \int_0^\infty \frac{d^3k}{(2\pi)^3}\left(1 - \frac{\left(k^2/2m_+ - \mu_s\right)}{\sqrt{\left(k^2/2m_+ - \mu_s\right)^2 + \Delta^2}}\right) \tag{B6}$$

Eq. (B6) can also be obtained directly from

$$n_s\left(T=0\right) = 2\int_0^\infty \frac{d^3k}{(2\pi)^3} v_k^2 \tag{B7}$$

## Appendix C: Grand canonical thermodynamic potential

### Superfluid case

After substituting the diagonalized Hamiltonian of the superfluid component, partition function becomes

$$Z = e^{-\frac{1}{T}\left(\sum_k\left(\left(k^2/2m_+ - \mu_s - \sqrt{\left(k^2/2m_+ - \mu_s\right)^2 + \Delta^2}\right) + \frac{\Delta^2}{V}\right)\right)} \prod_k\left(1 + e^{-E_{k_\alpha}/T}\right)\left(1 + e^{-E_{k_\beta}/T}\right) \tag{C1}$$

then, by using the relation between the grand-canonical thermodynamic potential for superfluid component, $\Omega_s$, and partition function, i.e., $\Omega_s = -T\ln Z$, then Eq. (C1) gives Eq. (23). At $T=0$, the terms proportional to $T\ln\left(1 + e^{-E/T}\right)$ vanish, and $\Omega_s$ reduces to

$$\Omega_s = \frac{\Delta^2}{g} + \sum_k\left(k^2/2m_+ - \mu_s - \sqrt{\left(k^2/2m_+ - \mu_s\right)^2 + \Delta^2}\right) \tag{C2}$$

Using the definition of the gap, the first term can be rewritten as

$$\Delta\frac{\Delta}{g}=\Delta\sum_{\vec{k}}u_{\vec{k}}v_{\vec{k}}=\sum_{\vec{k}}\frac{\Delta^2}{2\sqrt{\left(k^2/2m_+-\mu_s\right)^2+\Delta^2}} \tag{C3}$$

Finally, $\Omega_s$ becomes

$$\Omega_s=\sum_k\left(k^2/2m_+-\mu_s-\frac{\left(k^2/2m_+-\mu_s\right)^2+\left(\Delta^2/2\right)}{\sqrt{\left(k^2/2m_+-\mu_s\right)^2+\Delta^2}}\right) \tag{C4}$$

After using dimensional regularization[72] and Eqs. (B1), and converting the summation to an integral, Eq. (C4) reduces to Eq. (24).

**Normal case**

Noting that in the zero-temperature limit, the logarithm term in Eq. (27) vanishes when $k^2/2m_\sigma>\mu_\sigma-U_{N_\sigma}$ and reduces to $-(1/T)\left(\left(k^2/2m_\sigma\right)-\mu_\sigma-U_{N_\sigma}\right)$ when $k^2/2m_\sigma<\mu_\sigma-U_{N_\sigma}$, then, $\Omega_{N_\sigma}$ for each spin species becomes

$$\Omega_{N_\sigma}=\sum_{\vec{k}}\frac{k^2}{2m_\sigma}-\left(\mu_\sigma-U_{N_\sigma}\right)n_\sigma \tag{C5}$$

where $n_\sigma$ is the number of spin $\sigma$, and is given by $\int d^3k/(2\pi)^3$. After using transforming the summation to an integral, i.e., $\sum_{\vec{k}}\ldots\to\left(1/(2\pi)^3\right)\int d^3k\ldots$, finally, $\Omega_{N_\sigma}$ becomes

$$\Omega_{N_\sigma}=\left(\int_{k\le k_{F\sigma}=\sqrt{2m_\sigma\left(\mu_\sigma-U_{N\sigma}\right)}}\frac{d^3k}{(2\pi)^3}\left(\frac{k^2}{2m_\sigma}-\left(\mu_\sigma-U_{N_\sigma}\right)\right)\right)=\left(\frac{k_{F,\sigma}^5}{10\pi^2m_\sigma}-\left(\mu_\sigma-U_{N_\sigma}\right)\frac{k_{F,\sigma}^3}{6\pi^2}\right)\Bigg|_{k_{F\sigma}=\sqrt{2m_\sigma\left(\mu_\sigma-U_{N\sigma}\right)}} \tag{C6}$$

## References

[1] M.W. Zwierlein, A. Schirotzek, C.H. Schunck, W. Ketterle, Science 311, 492 (2006).
[2]G.B. Partridge, W. Li, R.I. Kamar, Y.A. Liao, R.G. Hulet, Science 311, 503 (2006).
[3] D. M. Eagles, Phys. Rev. 186, 456 (1969).
[4] A. J. Leggett, “Diatomic molecules and cooper pairs,” in Modern Trends in the Theory of Condensed Matter (Springer Berlin Heidelberg, Berlin, Heidelberg, 1980).
[5] P. Nozières and S. Schmitt-Rink, J. Low Temp. Phys. **59**, 195 (1985).
[6] C. A. R. Sá de Melo, M. Randeria, and J. R. Engelbrecht, Phys. Rev. Lett. **71**, 3202 (1993).
[7] J. R. Engelbrecht, M. Randeria, and C. A. R. Sá de Melo, Phys. Rev. B **55**, 15153 (1997).
[8] F. Pistolesi and G. C. Strinati, Phys. Rev. B **49**, 6356 (1994).

[9]M. Marini, F. Pistolesi and G.C. Strinati, Euro. Phys. J. B 1,151 (1998).
[10]N. Andrenacci, A. Perali, P. Pieri, and G.C. Strinati, Phys. Rev. B **60**, 12410 (1999).
[11] R. Haussmann, Phys. Rev. B **49**, 12975 (1994).
[12] Q. Chen, I. Kosztin, B. Jankó, and K. Levin, Phys. Rev. Lett. **81**, 4708 (1998).
[13] E. Tiesinga, B. J. Verhaar, and H. T. C. Stoof, Phys. Rev. A **47**, 4114 (1993).
[14] E. Timmermans, P. Tommasini, M. Hussein, and A. Kerman, Phys. Rep. **315**, 199 (1999).
[15]P.F.Bedaque, H. Caldas, G. Rupak, Phys. Rev. Lett.91, 247002 (2003).
[16]H. Caldas, Phys. Rev. A69, 063602 (2004).
[17] H. Feshbach, Ann. Phys.5, 357 (1958).
[18] K. Machida, T. Mizushima, and M. Ichioka, Phys. Rev. Lett. 97, 120407 (2006).
[19] M. Haque, H.T.C. Stoof, Phys. Rev. A74, 011602 (2006).
[20] T.N. De Silva, E.J. Mueller, Phys. Rev. A73, 051602 (2006).
[21] D.E. Sheehy, L. Radzihovsky, Ann. Phys. (NY)322, 1790 (2007).
[22] S. Giorgini, L. P. Pitaevskii, S. Stringari, Rev. Mod. Phys.80, 1215 (2008). L. P. Pitaevskii
[23] N.R. Cooper, Adv. Phys.57, 539 (2008).
[24] A.J. Leggett, Rev. Mod. Phys.47, 331 (1975).
[25] Q. Chen, J. Stajic, S. Tan, K. Levin, Phys. Rep. 412, 1 (2005).
[26] D. M. Bauer, M. Lettner, C. Vo, G. Rempe, S. Dürr, Nature Physics 5, 339 (2009).
[27] T. Mizushima, K. Machida, M. Ichioka, Phys. Rev. Lett.94, 060404 (2005).
[28] D. Inotani, R. Watanabe, M. Sigrist, and Y. Ohashi, Phys. Rev. A 85, 053628 (2012).
[29] T. Kashimura, R. Watanabe, and Y. Ohashi, Phys. Rev. A 89, 013618 (2014).
[30]M. Pini, P. Pieri, and G. C. Strinati, Phys. Rev. B 107, 054505 (2023)
[31]M. Pini, P. Pieri, R. Grimm, and G. C. Strinati, Phys. Rev. A 103, 023314 (2021).
[32]B. Van Schaeybroeck, A. Lazarides, Phys. Rev. A79, 053612 (2009).
[33]Y. Shin, M.W. Zwierlein, C.H. Schunck, A. Schirotzek, W. Ketterle, Phys. Rev. Lett. 97, 030401 (2006).
[34]A.M. Clogston, Phys. Rev. Lett.9, 266 (1962).
[35]B. S. Chandrasekhar, Appl. Phys. Lett. 1, 7 (1962).
[36]M. Iskin and C. A. R. S´a de Melo, Phys. Rev. Lett. **97**, 100404 (2006).
[37] M. M. Parish, F. M. Marchetti, A. Lamacraft, B. D. Simons, Phys. Rev. Lett. 98, 160402 ( 2007).
[38]J. E. Baarsma, K. B. Gubbels, and H. T. C. Stoof, Phys. Rev. A **82**, 013624 (2010).
[39]J. Wang, H. Guo, and Q. Chen, Phys. Rev. A 87, 041601(R) (2013).
[40]I. Bausmerth, A. Recati, and S. Stringari, Phys. Rev. A 79, 043622 (2009).
[41]S. S. Chung, C. J. Bolech, Phys. Rev. A, 96, 023609 (2017).
[42]E. Fratini and S. Pilati, Phys. Rev. A 90, 023605 (2014).
[43]G.-D. Lin, W. Yi and L.-M. Duan, Phys. Rev. A **74**, 031604(R) (2006).
[44] C. Ravensbergen, E. Soave, V. Corre, M. Kreyer, B. Huang, E. Kirilov, and R. Grimm, Phys. Rev. Lett. 124, 203402 (2020).
[45] E. Soave, A. Canali, Zhu-Xiong Ye, M. Kreyer, E. Kirilov, and R. Grimm, Phys. Rev. Research 5, 033117 (2023).
[46] Zhu-X. Ye, A. Canali, E. Soave, M. Kreyer, Y. Yudkin, C. Ravensbergen, E. Kirilov, and R. Grimm, Phys. Rev. A 106, 043314 (2022).
[47] M. J. H. Ku, A. T. Sommer, L. W. Cheuk, M. W. Zwierlein, Science 335. 563 (2012).
[48] P. van Wyk, H. Tajima, R. Hanai, Y. Ohashi, Phys. Rev. A 93, 013621(2016).
[49]S. K. Baur, S. Basu, T.N. De Silva, and E. J. Mueller, Phys. Rev. A **79**, 063628 (2009).
[50]X-J Liu, H. Hu, P. D. Drummond, Phys. Rev. A **75**, 023614 (2007).
[51]C-C. Chien, Q. Chen, Y. He, and K. Levin, Phys. Rev. Lett. **97**, 090402 (2006).
[52]M.M. Parish, S.K. Baur, E.J. Mueller, D.A. Huse, Phys. Rev. Lett. **99**, 250403 (2007).
[53]S.N. Klimin, J. Tempere, J.P.A. Devreese, J. Low Temp. Phys. 165, 261 (2011).
[54]Q. Chen, Y. He, C.C. Chien, K. Levin, Phys. Rev. B **75**, 014521 (2007).
[55]B. A. Olsen, M. C. Revelle, J. A. Fry, D. E. Sheehy, and R. G. Hulet, Phys. Rev. A 92, 063616 (2015).

[56]L. Radzihovsky and D. E Sheehy, *Rep. Prog. Phys.* **73,** 076501 (2010).
[57]D. E. Sheehy and L. Radzihovsky, Phys. Rev. Lett. **96**, 060401 (2006).
[58]F. Chevy, Phys. Rev. A **74**, 063628 (2006).
[59]F. Chevy and C. Mora, *Rep. Prog. Phys.* **73** 112401 (2010).
[60]T. Mizushima, M. Ichioka, K. Machida, J. Phys. Soc. Jpn. 76, 104006 (2007).
[61]H. Caldas, and A. L. Mota, *J. Stat. Mech.* 2008, 08013 (2008).
[62]P. Zdybel and P. Jakubczyk, *J. Phys.: Condens. Matter* **30** 305604 (2018).
[63]J. Carlson and S. Reddy, Phys. Rev. Lett. **95**, 060401 (2005).
[64]S. Pilati and S. Giorgini, Phys. Rev. Lett. **100**, 030401 (2008).
[65]Y. He, C.-C. Chien, Q. Chen, and K. Levin, Phys. Rev. B **76**, 224516 (2007).
[66]H. Guo, C.-C. Chien, Q. Chen, Y. He, and K. Levin, Phys. Rev. A **80**, 011601(R) (2009).
[67] B. Van Schaeybroeck, A. Lazarides, Phys. Rev. Lett. 98, 170402(2007).
[68]N. Ebrahimian, M. Mehrafarin, R. Afzali, Physica B407, 140 (2012).
[69]G. Bertaina and S. Giorgini ,Phys. Rev. A 79, 013616 (2009).
[70]A. Recati, C. Lobo, and S. Stringari, Phys. Rev. A 78, 023633 (2008).
[71] T. Papenbrock and G. F. Bertsch, Phys. Rev. C 59, 2052 (1999).
[72] J.C. Collins, Renormalization, Cambridge Univ. Press, Cambridge (1984).
[73] P-G de Gennes, Superconductivity of Metals and Alloys (Westview Press, USA, 1999).
[74] J. B. Ketterson, S. N. Song, Superconductivity (Cambridge University Press, Cambridge, 1999).
[75] K. B. Gubbels, M. W. J. Romans, and H. T. C. Stoof, Phys. Rev. Lett. **97**, 210402 (2006).
[76] J. Wang, Y. Che, L. Zhang, and Q. Chen, Scientific Reports **7**, 39783 (2017).
[77]N. Ebrahimian, Eur. Phys. J. Plus 140: 1066, 1 (2025).
[78] I.S. Gradshteyn and I.M. Ryzhik, Table of Integrals, Series and Products (1980).